\documentclass[longauth]{aa}
\usepackage{lscape}
\usepackage{graphicx,subcaption}
\usepackage{txfonts}
\usepackage{natbib}
\usepackage{tabularx}
\usepackage{amsmath}
\usepackage{diagbox}
\usepackage{enumitem}
    \makeatletter
    \def\@listiii{\leftmargin\leftmarginiii
                   \labelwidth\leftmarginiii
                   \advance\labelwidth-\labelsep
                   \topsep\z@
                   \parsep\z@
                   \partopsep\z@
                   \itemsep\topsep}
    \makeatother
    
\bibpunct{(}{)}{;}{a}{}{,}
\newcommand{\msun}{{M$_\odot$}}

\def\beq{\begin{equation}}
\def\eeq{\end{equation}}

\begin{document}

\title{The VANDELS ESO public spectroscopic survey:}
\subtitle{\bf final Data Release of 2087 spectra and spectroscopic measurements \thanks{This paper presenting the final data release of the last high redshift VIMOS survey is dedicated to the memory Olivier Le Fevre, PI of the VIMOS instrument, world known expert of extragalactic spectroscopy  and a pioneer in spectroscopy of the distant Universe.}}

\author{
B. Garilli \inst{1}
\and  R. McLure  \inst{2}
\and L. Pentericci  \inst{3}
\and P.  Franzetti \inst{1}
\and A. Gargiulo \inst{1}
\and A. Carnall \inst{2}
 \and O.   Cucciati   \inst{4}
 \and  A.   Iovino     \inst{5}
\and R.   Amorin  \inst{6,7}  
\and  M.   Bolzonella \inst{4}
\and  A.   Bongiorno   \inst{3}
\and  M.   Castellano  \inst{3}
\and  A.   Cimatti   \inst{8,9} 
\and  M.   Cirasuolo  \inst{10}
\and  F.   Cullen   \inst{2}  
\and  J.   Dunlop  \inst{2}   
\and  D.   Elbaz \inst{12}    
\and  S.   Finkelstein \inst{13}
\and  A.   Fontana   \inst{3}  
\and  F.   Fontanot  \inst{14,48}
\and  M.   Fumana  \inst{1}
\and  L.   Guaita   \inst{15} 
\and  W.   Hartley  \inst{16}
\and  M.   Jarvis     \inst{17}
\and  S.   Juneau     \inst{54}
\and D.   Maccagni  \inst{1}
\and  D.   McLeod  \inst{2}
\and  K.   Nandra     \inst{18}
\and  E.   Pompei     \inst{19}
\and  L.   Pozzetti   \inst{4}
\and  M. Scodeggio \inst{1}
\and  M. Talia  \inst{8,4}
 \and A. Calabro'  \inst{3}
 \and  G. Cresci  \inst{9}
\and   J. P.U. Fynbo  \inst{20}
\and  N. P. Hathi  \inst{21}
\and P. Hibon \inst{19}
\and A. M. Koekemoer\inst{21}
\and   M. Magliocchetti  \inst{22}
\and    M. Salvato  \inst{18}
\and   G. Vietri \inst{1}
\and    G. Zamorani  \inst{4}
\and O. Almaini\inst{23}
\and I. Balestra\inst{24}
\and S. Bardelli\inst{5}
\and R.Begley\inst{2}
\and G. Brammer\inst{20}
\and E. F. Bell\inst{25}
\and R.A.A. Bowler\inst{17}
\and M. Brusa\inst{8}
\and F. Buitrago\inst{26,27,49}
\and C. Caputi\inst{28}
\and P. Cassata\inst{29}
\and S. Charlot\inst{30}
\and A. Citro\inst{4}
\and S. Cristiani\inst{14}
\and E. Curtis-Lake\inst{30}
\and M. Dickinson\inst{31}
\and G. Fazio\inst{32}
\and H.C. Ferguson\inst{33}
\and F. Fiore\inst{14}
\and M. Franco\inst{12, 52}
\and A. Georgakakis\inst{18}
\and M. Giavalisco\inst{34}
\and A. Grazian\inst{35}
\and M. Hamadouche\inst{2}
\and I. Jung\inst{50, 51}
\and S. Kim\inst{36}
\and Y. Khusanova\inst{37}
\and O. Le F\`evre\inst{37}
\and M. Longhetti\inst{5}
\and J. Lotz\inst{33}
\and F. Mannucci\inst{9}
\and D. Maltby\inst{23}
\and K. Matsuoka\inst{9}
\and H. Mendez-Hernandez\inst{38}
\and J. Mendez-Abreu\inst{39,40}
\and M. Mignoli\inst{4}
\and M. Moresco\inst{4,8}
\and M. Nonino\inst{14}
\and M. Pannella\inst{41}
\and C. Papovich\inst{42}
\and P. Popesso\inst{43}
\and G. Roberts-Borsani\inst{53}
\and D.J. Rosario\inst{44}
\and A. Saldana-Lopez\inst{11}
\and P. Santini\inst{3}
\and A. Saxena\inst{16}
\and D. Schaerer\inst{11}
\and C. Schreiber\inst{45}
\and D. Stark\inst{46}
\and L.A.M. Tasca\inst{37}
\and R. Thomas\inst{19}
\and E. Vanzella\inst{4}
\and V. Wild\inst{47}
\and C. Williams\inst{46}
\and E. Zucca\inst{4}
}
   
\institute{
INAF - Istituto di Astrofisica Spaziale e Fisica Cosmica Milano, via A.Corti 12,
20133 Milano, Italy
\and
Institute for Astronomy, University of Edinburgh, Royal Observatory, Edinburgh, EH9 3HJ, UK
\and
INAF-Osservatorio Astronomico di Roma , via Frascati 33, I-00078 Monteporzio Catone, Italy 
\and
INAF-Osservatorio di Astrofisica e Scienza dello Spazio di Bologna, via Gobetti 93/3, I-40129, Bologna, Italy
\and 
INAF-Osservatorio Astronomico di Brera, via Brera 28, I-20122 Milano, Italy
\and Instituto de Investigaci\'on Multidisciplinar en Ciencia y Tecnolog\'ia, Universidad de La Serena, Ra\'ul Bitr\'an 1305, 
La Serena, Chile
\and Departamento de Astronom\'ia, Universidad de La Serena, Av. Juan Cisternas 1200 Norte,  La Serena, Chile
\and University of Bologna, Department of Physics and Astronomy ``Augusto Righi'' (DIFA), Via Gobetti 93/2, I-40129, Bologna, Italy 
\and INAF - Osservatorio Astrofisico di Arcetri, Largo E. Fermi 5, I-50157 Firenze, Italy
\and European Southern Observatory, Karl-Schwarzschild-Str. 2, D-85748 Garching b. Munchen, Germany
\and Observatoire de Gen\'eve, Universit\'e de Gen\'eve, 51 Ch. des Maillettes, 1290, Versoix, Switzerland 
\and Laboratoire AIM-Paris-Saclay, CEA/DSM/Irfu, CNRS France 
\and Department of Astronomy, The University of Texas at Austin, Austin, TX 78712, USA
\and INAF - Astronomical Observatory of Trieste, via G.B. Tiepolo 11, I-34143 Trieste, Italy
\and  N\'ucleo  de  Astronom\'ia,  Facultad  de  Ingenier\'ia,  Universidad  Diego Portales,  Av.    Ej\'ercito  441,  Santiago,  Chile
\and Department of Physics and Astronomy, University College London, Gower Street, London WC1E 6BT, UK
\and Astrophysics, The Denys Wilkinson Building, University of Oxford, Keble Road, Oxford, OX1 3RH
\and Max  Planck Institut f\"ur Extraterrestrische Physik Giessenbachstrasse 1, Garching D-85748, Germany
\and European Southern Observatory, Avenida Alonso de C\'ordova 3107, Vitacura, 19001 Casilla, Santiago de Chile, Chile
\and The Cosmic Dawn Center, Niels Bohr Institute, University of Copenhagen, Juliane Maries Vej 30, 2100 Copenhagen, Denmark
\and Space Telescope Science Institute, 3700 San Martin Drive, Baltimore, MD, 21218, USA
\and INAF- Istituto di Astrofisica e Planetologia Spaziali,  Via del Fosso del Cavaliere 100, 00133 Roma, Italy
\and School of Physics and Astronomy, University of Nottingham, University Park, Nottingham NG7 2RD, UK 
\and University Observatory Munich, Scheinerstrasse 1, D-81679 Munich, Germany
\and Department of Astronomy, University of Michigan, 311 West Hall, 1085 South University Ave., Ann Arbor, MI 48109-1107, USA
\and Instituto de Astrof\'isica e Ci\^encias do Espa\c co, Universidade de Lisboa, OAL, Tapada da Ajuda, P-1349-018 Lisbon, Portugal
\and Departamento de F\'isica, Faculdade de Ci\^encias, Universidade de Lisboa, Edif\'icio C8, Campo Grande, PT1749-016 Lisbon, Portugal
\and Kapteyn Astronomical Institute, University of Groningen, Postbus 800, 9700 AV, Groningen, The Netherlands
\and Dipartimento di Fisica e Astronomia "Galileo Galilei" - DFA, vicolo dell' Osservatorio, 3, Padova, Italy 
\and Institute d'Astrophysique de Paris, CNRS, Universit\'e Pierre et Marie Curie, 98 bis Boulevard Arago, 75014, Paris, France
\and National Optical Astronomy Observatory, 950 North Cherry Ave, Tucson, AZ, 85719, USA
\and Harvard-Smithsonian Center for Astrophysics, 60 Garden St, Cambridge MA 20138
\and  Dark Cosmology Centre, Niels Bohr Institute, University of Copenhagen, Juliane Maries Vej 30, DK-2100 Copenhagen, Denmark
\and Astronomy Department, University of Massachusetts, Amherst, MA 01003, USA 
\and INAF- Osservatorio Astronomico di Padova,  Vicolo dell'Osservatorio, 5, 35122 Padova, Italy 
\and Pontificia Universidad Cat\'olica de Chile  Instituto de Astrof\'isica Avda. Vicu\~na Mackenna 4860 - Santiago - Chile 
\and Aix Marseille Universit\'e, CNRS, LAM (Laboratoire d'Astrophysique  de Marseille) UMR 7326, 13388, Marseille, France 
\and Instituto de Fisica y Astronomia, Facultad de Ciencias, Universidad de Valparaiso, 1111 Gran Bretana, Valparaiso, Chile
\and Instituto de Astrof\'isica de Canarias, Calle Via L\'actea s/n, E-38205 La Laguna, Tenerife, Spain
\and Departamento de Astrof\'isica, Universidad de La Laguna, E-38200 La Laguna, Tenerife, Spain
\and Faculty of Physics, Ludwig-Maximilians Universit\"at, Scheinerstr. 1, 81679, Munich, Germany
\and Department of Physics and Astronomy, Texas A\&M University, College Station, TX 77843-4242, USA
\and Excellence Cluster, Boltzmannstr. 2, D-85748 Garching, Germany
\and Department of Physics, Durham University, South Road, DH1 3LE Durham, UK
\and Leiden Observatory, Leiden University, 2300 RA, Leiden, The Netherlands 
\and Steward Observatory, The University of Arizona, 933 N Cherry Ave, Tucson, AZ, 85721, USA
\and School of Physics and Astronomy, University of St. Andrews, SUPA, North Haugh, KY16 9SS St. Andrews, UK
\and IFPU - Institute for Fundamental Physics of the Universe, via Beirut 2, 34151, Trieste, Italy
\and Departamento de F\'{i}sica Te\'{o}rica, At\'{o}mica y \'{O}ptica, Universidad de Valladolid, 47011 Valladolid, Spain
\and Astrophysics Science Division, Goddard Space Flight Center, Greenbelt, MD 20771, USA
\and Department of Physics, The Catholic University of America, Washington, DC 20064, USA
\and Centre for Astrophysics Research, School of Physics, Astronomy \& Mathematics, University of Hertfordshire, College Lane, Hatfield AL10 9AB, UK
\and Department of Physics and Astronomy, PAB, 430 Portola Plaza, Box 951547, Los Angeles, CA 90095-1547, USA
\and NSF's NOIRLab, 950 N Cherry Ave, Tucson, AZ 85719, USA
}

\offprints{B.Garilli, bianca.garilli@inaf.it}

\date{Received 4 December 2020; accepted 18 January 2021}

\abstract 
{VANDELS is an ESO Public Spectroscopic Survey 
 designed to build a  sample of high signal to noise, medium resolution spectra of galaxies at redshift between 1 and 6.5.
 Here we present the final  Public Data Release of the VANDELS  Survey,
comprising 2087 
redshift  measurements.
We  give 
a detailed description of sample selection, 
observations and data reduction procedures. The final catalogue reaches  a target selection completeness of 40\% at $i\mathrm{_{AB}=25}$. The high Signal to Noise ratio of the spectra (above 7 in 80\% of the spectra) and the dispersion of 2.5\AA~allowed us to measure redshifts with high precision, the redshift measurement success rate reaching almost 100\%. Together with the redshift catalogue and the reduced spectra, we  also provide  optical mid-IR photometry and  physical parameters derived through SED fitting. The observed galaxy sample comprises both passive and star forming galaxies covering a stellar mass range 8.3< Log(M$_{*}$/M$_\odot$)<11.7. 
All catalogues and spectra are accessible  through the survey
database ({\tt http://vandels.inaf.it}) where all information can be
queried interactively, and via the ESO Archive ({\tt https://www.eso.org/qi/}).} 

 
   


\authorrunning{B.Garilli et al.}
\titlerunning{VANDELS- Final Public Data Release}
\maketitle

\section{Introduction}
Understanding when and how galaxies formed from the first gas clouds, and  evolved to their variety of morphologies and properties as observed in the local universe,  is one of the key questions   of extragalactic astrophysics, which presents both theoretical and observational challenges. 
With the advent of multi-object spectrographs mounted on 10-m class telescopes,  
spectroscopic surveys of distant galaxies have entered the epoch of statistical studies. Starting with the local Universe in the late 1990s with the 2DF \citep{2df} and SDSS  surveys (from DR1, \citealt{sdss1}, to DR16, \citealt{sdss16}), the exploration of the statistical properties of galaxies moved further and further in redshift: among others we remember the pioneering works of CFRS  \citep{cfrs} and ESP \citep{esp}, VVDS \citep{vvds, vvds_wide}, DEEP \citep{deep}, zCosmos \citep{zcosmos_main} and VIPERS \citep{vipers_main,vipersfinal} at  <z>$\sim$0.7,
which have been able to collect some tens of thousands of redshifts, KBSS-MOSFIRE \citep{kbss_mosfire} and VUDS \citep{VUDS}, which have succeeded in collecting a few thousand  redshifts  between z=2 and z=6, till the smaller samples at very high redshift 
\citep{steidel2003,steidel2004,VLT_LBG,candelsz7,muse,kmos}. In parallel to surveys based upon optically selected samples, smaller surveys based on K-selected samples have been carried out (e.g. K20, \citealt{k20}, GMASS, \citealt{gmass}). Although spectroscopy was done in the optical range, for lack of Multi Object spectrograph operating in the Near Infrared, these  works have allowed to gain a different view of the galaxy population at medium-high redshifts. The last survey along this line is LEGA-C \citep{legac}, the other ESO public spectroscopic survey carried out in parallel to VANDELS.
These surveys all aimed at making a census of the galaxy population in the targeted redshift ranges, and they have allowed us a number of steps forward in our understanding of the evolution of galaxies. Thanks to the large statistics accumulated, luminosity and mass functions, correlation functions, the influence of the environment, and (in a lesser measure)  the  mass-metallicity relation and the mass-SFR relations are well known in the local Universe and up to z$\sim$1, while for the most directly observable relations (like the luminosity and mass functions) we have a good knowledge till redshift about 7. Still, due to the interplay between the quantities involved, more sophisticated diagnostics like Star Formation, metallicity and internal dust absorption still suffer from large uncertainties and no clear discrimination can yet be made among the different evolutionary scenarios. 

VANDELS,  proposed as an ESO public spectroscopic survey in 2014, aims at throwing new light on these aspects, not limiting itself to finding a redshift, but also at providing mid resolution, high signal to noise (S/N) spectra which allow to study in detail and with  statistically meaningful numbers the physical characteristics of the high redshift galaxies \citep{mclure2018}. Since the first data release of VANDELS \citep{vandelsdr1}, a number of different studies have already been published: from  dust attenuation and stellar metallicities of star forming  \citep{cullen2018,cullen2019,calabro} and quiescent galaxies \citep{carnall2019,carnall2020}, to Ly$\alpha$ and He II~$\lambda$1640 
emitters \citep{marchi2019,hoag2019,cullen2020,saxena2020a,saxena2020b,guaita} to Intergalactic medium properties 
\citep{Thomas2020} and AGNs \citep{maglioc2020}. 
All these works were based on only a subset of the data. In this paper we present the full VANDELS data set which is being released to the whole astronomical community, complete with redshifts, spectra and SED-fitting derived physical properties, and give all the information required to fully exploit the scientific content of the VANDELS data-set.

The layout of the paper is as follows: \S2
summarises the survey strategy and design;
\S3 describes the VLT-VIMOS observations; \S4 discusses
the data reduction, including redshift measurement and description of the redshift quality flags;
\S5 presents the VANDELS final sample, discussing redshift errors and
comparison to external data; 
\S6
provides examples of VANDELS spectra; 
\S7  discusses the SED-fitting derived quantities for the spectroscopic sample, and presents the main relations of the spectroscopic sample compared to the parent photometric sample; 
\S8 provides information on the access to the VANDELS data set;
finally, \S9 provides a brief summary.
Throughout this paper, we use a Concordance Cosmology with
$\Omega_{\rm m}=0.3$, $\Omega_{\Lambda}=0.7$ and 
$H_{0}=70~\mathrm{km~s^{-1}~Mpc^{-1}}$ and
adopt a \citet{chabrier} initial mass function (IMF) for calculating
stellar masses and SFRs.. 
\\
\section{Survey strategy and design}
\label{design}
\subsection{Photometric Catalog}
\label{parent}
VANDELS is an extragalactic ESO Public Spectroscopic Survey carried out using the VIMOS spectrograph \citep{vvds_tech} on the VLT. It has been designed 
to obtain ultra-deep medium resolution spectra with S/N high enough to allow measurement of spectral lines from the individual brighter sources, or from stacked spectra of the fainter ones. 
VANDELS targets two separate survey fields, UDS 
(Ultra Deep Survey, RA=2:18m,  DEC=-5:10) and CDFS (Chandra Deep Field South, RA=3:32, DEC=-27:48),
covered by different sets of imaging data, thus  target
selection had to be performed using four independent
photometric catalogues. Furthermore, within each field the footprint of VIMOS is such that we had to extend the central part, covered by HST, with an external part covered only by ground-based photometry
(see Figure~\ref{pointings}).
As a result, the VANDELS
survey started from 4 different photometric catalogues: 
UDS-HST,
UDS-GROUND, CDFS-HST and CDFS-GROUND.  
More details on the catalogues, as well as on the computation of photometric redshifts at the base of our selection and on the target selection itself, are given in \citet{mclure2018}, and summarised here.

Within the two regions covered by the WFC3/IR imaging provided
by the CANDELS survey \citep{candels1,candels2} (UDS-HST and CDFS-HST), we used
the H-band-selected photometric catalogues produced by the CANDELS
team \citep{Galametz, Guo}. 
Within the wider-field areas, at the time of the survey design 
near-IR-selected photometric catalogues meeting 
the magnitude limit we impose for target selection (see Section \ref{targetDef})
were not publicly available. As a result, new multi wavelength photometric
catalogues were generated using the publicly available imaging, covering 12 (17) bands in UDS (CDFS). 
Object detection was performed in H-band and photometry  measured within 2" diameter circular
apertures. The depth of the UDS (CDFS) ground-based catalogues reaches mag~27 (26.5) in the optical bands, and mag~25 (24.5) in the NIR-bands.

VANDELS targets are pre-selected on the basis of photometric redshifts. 
Having four  photometric catalogues, each comprising different bands at different depths, it was important to ensure homogeneity in the quality of the photometric redshifts. For the 2 areas 
 covered by deep HST near-IR
imaging (UDS-HST and CDFS-HST), we adopted the photometric
redshifts made publicly available by the CANDELS survey team
\citep{Santini}. 
For the wider area regions new photometric redshifts based on the new UDS-GROUND and CDFS-GROUND
photometric catalogues were computed  within
the VANDELS team by taking the median value of 14 different estimates derived
by different members of the team using different public and private  codes. 
Comparing these values with the ones provided by the CANDELS survey team 
using various spectroscopic validation
sets, \citet{mclure2018} quantify the accuracy of the final
photometric redshifts adopted for the wider area UDS-GROUND
and CDFS-GROUND regions as $\mathrm{\sigma_{dz}=0.017}$ with an outlier rate of 1.9\%, comparable to the accuracy obtained for the HST catalogues.

Finally,  in order to produce the cleanest possible selection catalogue, it
was necessary to remove potential stellar sources. Within 
the UDS-HST and
CDFS-HST regions we excluded all sources having a SEXTRACTOR \citep{sextractor} stellarity parameter
CLASS\_STAR $\geq$ 0.98 in the \citet{Galametz} and \citet{Guo} catalogues . 
For the two ground-based photometric catalogues, we excluded
as stars all sources
consistent with the stellar locus on the  BzK diagram by \citet{Daddi04}.
Secondly, we performed an SED fitting of all remaining sources using a range of stellar templates drawn from the SpeX
archive \citep{Burgasser}, and removed all sources which produced an improved
SED fit with a stellar template and were consistent with being a point
source at ground-based resolution. Indeed, within our measured sample only one object turned out to be a star.

Using such clean and deep multi-wavelength photometric catalogues, with associated photometric redshifts, we  performed  SED fitting to derive 
SFRs, stellar masses, and rest-frame photometry, and we based our source classification on the basis of these SED derived physical properties. We defined as star-forming galaxies objects having $\mathrm{sSFR > 0.1 Gyr^{-1}}$ at z$_{phot}$>2.4, as {\it passive} galaxies objects in the redshift range 1<z$_{phot}$<2.5 satisfying the colour-colour criteria \citep{williams}: U - V  > 0.88(V - J ) + 0.49,
U - V > 1.2,
V - J < 1.6. Among star forming galaxies, we have further defined as Lyman break those galaxies within the 
redshift range $3.0 \leq \mathrm{z_{phot}} \leq 7.0$

\subsection{Target definition}
\label{targetDef}
The VANDELS spectroscopic targets
were all pre-selected using the high-quality photometric redshifts and the classification described above, 
with the vast majority ($\sim 97\%$) drawn from three main categories (see Table \ref{surveyTable}, column 3).

\begin{enumerate}
\item Bright ($\mathrm{i_{AB} \le }$25) star-forming galaxies
in the redshift range $2.4 \le \mathrm{z_{phot}} \le 5.5 $. For these
galaxies, the aim was to get spectra with  
S/N per resolution element  larger than 10 
to allow stellar metallicity
and gas outflow information to be extracted from the individual objects (from here on the {\it star forming (SF)} sample)
\item A sample of massive
($\mathrm{H_{AB} \le 22.5 }$) passive galaxies at $\mathrm{1.0 \le \mathrm{z_{phot}} \le 2.5}$. In combination with deep multi-wavelength photometry
and 3D-HST grism spectroscopy \citep{brammer2012} the 
high S/N spectra 
(at least 10 per resolution element)
obtained by VANDELS are designed to provide
age/metallicity information and star-formation history constraints
for individual objects (from here on the {\it passive} sample)
\item  A large statistical sample
of faint star-forming galaxies ($\mathrm{25 \le H_{AB} \le 27}$, $i\mathrm{_{AB} \le 27.5}$) in
the redshift range $\mathrm{3 \le \mathrm{z_{phot}} \le 7}$ (median $\mathrm{z_{phot} = 3.5}$) (from here on the {\it Lyman break galaxies sample, LBG}, though they have not been selected using the classical colour-colour criterion). 
For this sample the main goal is redshift measurement, thus we require a S/N per resolution element larger than 5.
\end{enumerate}

To these three main categories, we have added a small sample of AGN candidates and Herschel detected sources. 
The AGNs candidates  all lie within the CDFS field and were selected
based on either a power-law SED shape in the mid-IR \citep{chang} or X-ray emission \citep{xue, rangel, hsu}.
We have restricted our selection to AGNs with $\mathrm{z_{phot}}\ge2.4$ and $i \le 27.5$ if within CDFS-HST, or $i \le 24$ if within CDFS-GROUND. We note here that the
photometric redshifts derived for the AGN candidates are based
on SED fitting with the same set of galaxy templates discussed in 
Section \ref{parent}, and are therefore not expected to be as accurate as the
photometric redshifts derived for the rest of the VANDELS sample. The Herschel
detected sources  lie either within the UDS-HST or the CDFS-HST regions,
have $\mathrm{z_{phot}} \ge 2.4$ and $i\mathrm{_{AB}} \le 27.5$, and are detected in at least one
Herschel band \citep{pannella}. From now on, we will name the spectroscopic {\it AGN} sample the ensemble of both  AGN candidates and Herschel sources.

\begin{figure*}
\resizebox{\hsize}{!}{\includegraphics[clip=true]{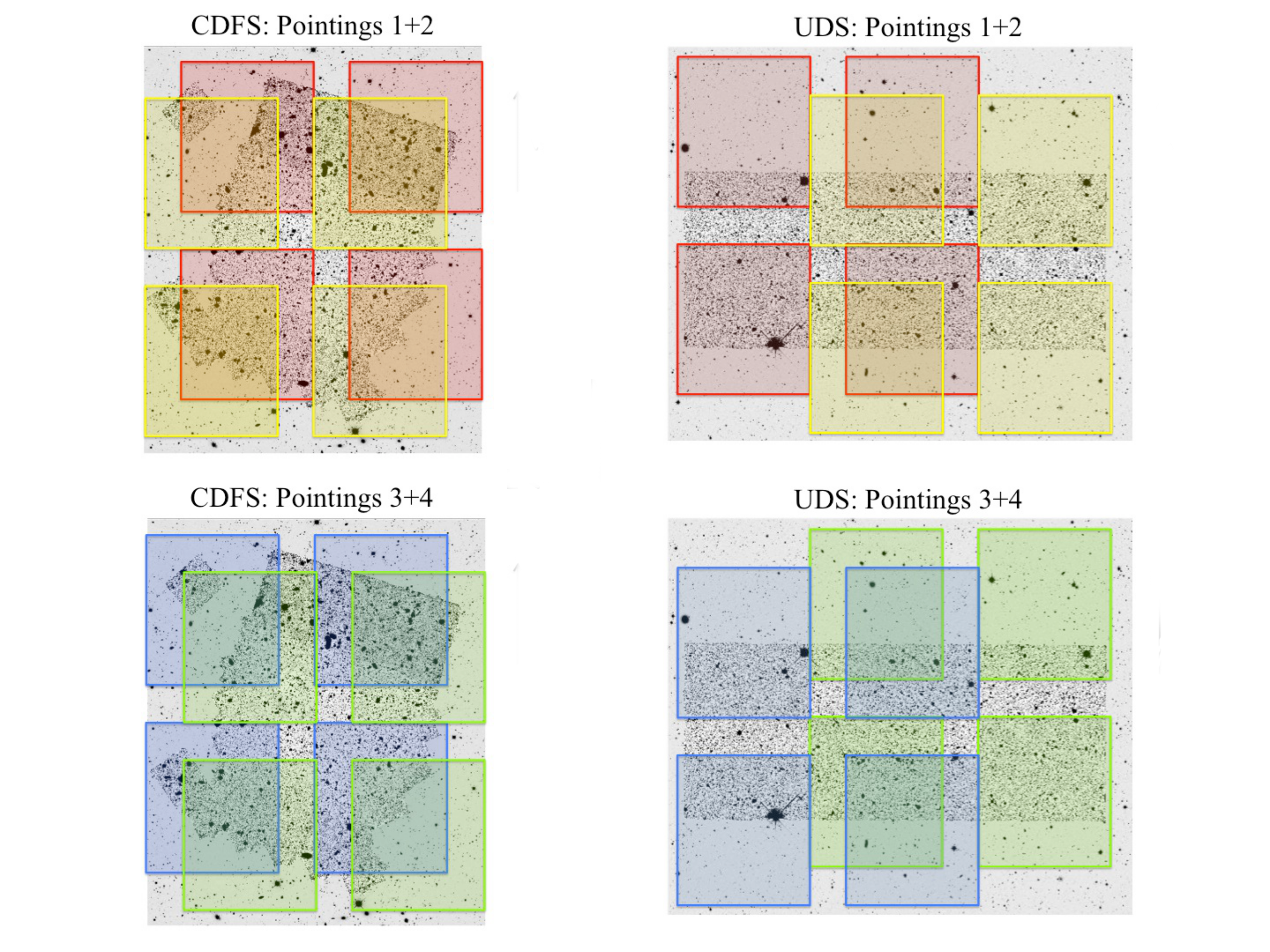} }
\caption{Layout of the VANDELS pointings, from \citet{mclure2018} Figure 1. Left CDFS and right UDS, North is up and East to the left. Each area is shown twice to better show the layout of the 4 VIMOS pointings (coloured squares). The grey scale image shows  the HST H-band imaging provided by the CANDELS survey \citep{candels1,candels2} in the central regions and the ground-based H-band imaging from  UKIDSS UDS (Almaini et al., in preparation) and VISTA VIDEO \citep{video} in the wider region.}
\label{pointings}
\end{figure*}

The exact layout of the 4 VANDELS pointings is shown in Figure~\ref{pointings}. The exact coordinates of the different VIMOS pointings (red/yellow and blue/green areas) have been chosen  to maximise the coverage of the area covered by HST photometry (darker areas in  Figure~\ref{pointings}). Table~\ref{surveyTable}, column 3, reports the total number of objects for each subsample.
\\
\subsection{Mask Design}
\label{maskdes}
The standard VIMOS observing procedure requires the acquisition of a direct image, which
is used for mask preparation with the {\sc vmmps}  software \citep{vmmps}
distributed by ESO: 
{\sc vmmps}  assigns the slit length taking
into account object dimensions
and sky subtraction regions as specified by the
user. While the slit width is set by the user, the slit  length is accomodated by the software as part of the
optimisation process taking into account user given minimal constraints, to maximise the number of slits per quadrant while ensuring alignment of spectra along the dispersion direction.

The target allocation for the full survey has been done once and forever at the beginning of the survey.
In order to fulfil the S/N requirements on the continuum  (for {\it passive} and bright {\it SF} galaxies), or on the emission lines (for the faint {\it SF} galaxies),  we adopted a nested slit
allocation strategy: within a given pointing, the brightest objects  appear on a single mask (and are observed for 20 hours),
fainter objects appear on two masks (40 hours exposure time) and the faintest objects appear on four masks (80 hours exposure time).  We carried out extensive simulations on the best slit allocation strategy which could maximise the total number of observed targets, while allowing a statistically significant sample  to be observed even for the lower surface density  sources (namely bright star-forming galaxies
and massive {\it passive} galaxies). From these simulations we have imposed the additional constraint of having approximately a 1:2:1 ratio for objects requiring 20:40:80 hours of integration time. 
No other additional prioritisation
(e.g. in terms of redshift or source brightness) was applied during the slit
allocation process.  
To ensure optimal sky background subtraction we adopted a 'nod along the slit' observing strategy, and imposed a minimum distance of 8 pixels (1.64 arcsec) between the source and the slit edge. Targets were treated as point-like sources, and a minimum slit length of 28 pixels (5.7 arcsec) was imposed. On average, it was possible to place  $\sim$50 slits per quadrant, and the final sample of observed targets is reported in Table~\ref{surveyTable}, column 4. 
\begin{table*}
\caption{VANDELS observed sample}
\label{surveyTable}
\centering
\begin{tabular}{c c c c c c }
\hline\hline
Field       & sample  &Potential  & Observed   &  Measured  & secure \\
             &                 &    targets &  targets  &   redshifts       & redshifts \\
\hline     
CDFS &   passive   & 307              & 124        &       123         & 122          \\
CDFS &   SF           & 745             & 201        &        201        &   196        \\
CDFS &   LBG        &  3277            & 626        &        604        &    514        \\
CDFS &   AGN       & 151            & 55           &        47       &    20       \\
\hline
CDFS & total          & 4480              & 1006     &       975         & 852    \\
CDFS & secondary obj. &                    &     55         &      44                &     28      \\
\hline
\hline
UDS &   passive    &505               &  157          &       155         & 153         \\
UDS &   SF           & 998              & 216       &       216         & 212         \\
UDS &   LBG        & 3645              & 672      &       655         & 540          \\
UDS &   AGN         & 28            & 10       &       10         & 2          \\
\hline
UDS & total           & 5176            & 1055   &       1036         & 907         \\
UDS & secondary obj. &                    &       49       &       32               &     24      \\
\hline
\hline
all targets & total &     9656          & 2061 &       2011        &  1759 \\
all secondary obj  & total &                &     104     &       76                &      52     \\
\hline
\end{tabular}
\end{table*}
\\
It is worth to perform an {\it a posteriori} check on whether the selected targets are a fair subsample of the full parent catalog. Using the Kolmogorov-Smirnov test we have tested whether the distributions of stellar mass and Star Formation Rate (SFR) as derived from the SED fitting for  the observed subsamples of {\it passive}, {\it SF} and {\it LBG }galaxies are drawn from the same parent population as the potential targets. For  the {\it passive} subsample, there is no indication of difference either in stellar mass or SFR distribution.  For the {\it SF} sample, the two samples of observed and parent catalogue galaxies are statistically identical within 3$\sigma$ if we limit  the comparison to $i\mathrm{_{AB}<24.5}$, while including the last half magnitude bin the two samples are statistically compatible only at 2$\sigma$, indicating that we start to loose low mass, low SFR galaxies in the last half magnitude bin.  This is expected, given that the long exposure times required by the  faintest objects are disfavoured by our allocation strategy 1:2:1.  For {\it LBGs}, the same considerations apply: they are the faintest objects and the sampling is the lowest. Restricting the comparison to $i\mathrm{_{AB}<26}$, the null hypothesis of the observed and parent sample being drawn from the same parent distribution, in terms of Mass and SFR, is valid at 2 $\mathrm{\sigma}$ level. Analogously to the {\it SF} galaxies, the observed {\it LBG} sample shows a loss of the low end of the Mass or SFR distributions. 
\\
\section{VLT-VIMOS Observations}
\label{observations}
In order to fulfil our scientific goals, we have used the 
`Medium Resolution' (MR) grism. Coupled with the  1-arcsec wide slits, a value which well matches
the average seeing in Paranal, the MR grism provides a spectral
resolution $\mathrm{R\simeq 650}$ and a mean dispersion of 2.5\AA/pixel in the wavelength range 4800-9800 \AA. All observations have been carried out in visitor mode, the established standard for ESO Public Surveys, between August 2015 and 
January 2018, after which VIMOS was decommissioned, with an average of 6 runs per year. Each single exposure was 20 minutes long, grouped by 3 in a standard Observation Block (OB) of 1 hour. Most observations were carried out with no moon: in those few cases when the moonlight illumination was higher than 30\%, distance from the moon was higher than 90 degrees. 75\% of the observations have an average airmass less than 1.2, and 92\% less than 1.4. Seeing, as measured directly on the science images, was below 1 arcsec in almost 90\% of the observations.
\\
Calibration exposures (flat fields and arc lines) were performed immediately before or after a 60 minutes scientific OB (i.e. every 2 hours), maintaining the instrument at the same rotation angle and inserting a screen at the Nasmyth focus. This ensures that we have
calibration lamps with the same flexure-induced distortions as the scientific
images, thus allowing for a more precise wavelength calibration. In order to minimise spectra distortions due to atmospheric refraction, observations were carried out aligning slits along the East-West direction and were  confined to within $\mathrm{\pm2}$ hours from
the meridian (see e.g. \citet{sanchez}).   
\\
\section{Data reduction} 
\label{reduction}
Data reduction
was performed using the recipes provided by the {\sc vipgi} package \citep{vipgi} and the {\sc easylife} environment \citep{easylife} already used for the VIPERS survey \citep{vipers_dr1}, adapted 
with a fully automated pipeline tailored for observations made across different nights and observing runs.
We summarise here the main concepts.

As a first step, in each raw frame the 2D dispersed 
spectra are located and 
traced. Each raw spectrum is collapsed along the dispersion direction, and the
object location computed.  A first sky subtraction is performed row by row,
avoiding the region identified as the object.
An inverse dispersion solution is computed for each column of each dispersed
spectrum making use of an arc calibration lamp. The wavelength calibration
uncertainty is always below  0.4\AA ~(1/6 of a pixel).
The inverse dispersion solution is applied before extraction.
A further check on the wavelength of sky lines is computed  on the linearized 2D
spectra, and, if needed, a rigid offset is applied to the data in order to bring the sky
lines to their correct wavelength. The ~60  scientific 20 minutes exposures of a 20 hours observation of the same
field are registered and co-added, and a second background subtraction is
performed repeating the procedure carried out before. Finally, 1D spectra are
extracted applying the Horne {\it optimal extraction} algorithm \citep{horne}, and spectra are 
corrected for the instrument sensitivity function, as derived from the standard
spectrophotometric observations routinely carried out by ESO. 
\\
Whenever a target required longer than 20 hours observing  time, it appeared in different masks. In these cases, instead of attempting extraction from the single 20-hour observations, where the signal to noise of the object was at, or below, the $3\sigma$ detection limit, we have preferred to  combine the 2D wavelength calibrated spectrograms (wavelength calibrated 2D spectra) of the single slits originating from each 20 minutes exposure, and perform the extraction on such deep spectrogram combinations. This ensures that we optimise the total signal to noise. The exposure to exposure offsets within the 20 hour subsets can be carefully computed using the brightest objects in the field   (namely those used at acquisition time to precisely align the mask), while the pointing differences between the 20 hour observations can be computed if the object is at least detected at  $1.5\sigma$ level in all the 20 hour subsets, a detection level which is reached for all objects. 
\\
The 1D spectrograms were corrected for the instrument sensitivity function using spectrophotometric standard stars. As the 1D spectra were extracted from the combination of a number of single exposures obtained over several observing runs, this operation allows only to correct for the instrument signature, i.e. to go from counts  to pseudo flux units. Absolute flux calibration was performed on a spectrum by spectrum basis, normalizing the spectrum to the available high quality photometry.
\\
\subsection{Blue end correction}
\label{bluecorr}

The very low instrument response below 5000\AA~and the use of late type stars as spectrophotometric standards, which optimises the measurement of the sensitivity function in the redder part of the spectrum, affects the precise computation of the sensitivity function in the bluest wavelength range. This had already been noted 
during final testing of the flux calibration of the DR1 spectra, comparing the spectra with the available photometry. Following an approach similar to the one used for DR1, 
we have implemented an
empirically derived correction to the spectra at these blue wavelengths
which accounts for the average flux loss. 
To compute the correction, we have used all flux calibrated spectra of galaxies in the redshift range 2.17 < z < 2.95 with the highest redshift quality flags (see \ref{redshift}). 
Such spectra should display a power-law continuum slope in the rest-frame wavelength
range ($1300 \AA \le \lambda \le 2400\AA$), as also confirmed by the available photometry.
After visual inspection we have discarded a small number of spectra which had obvious data reduction issues in the wavelength range of interest. The resulting 165 spectra were normalized by their median flux in the wavelength range $5750\AA< \lambda <8000\AA$~ and an 
observed-frame median stack was produced. The stack has been fitted with a power-law in the same  wavelength range and such fit has been extrapolated down to 4800\AA. The flux correction has been computed as 
the ratio of the power-law continuum to the stacked spectrum fitted with  a 5th order
polynomial. We have repeated the procedure keeping separated objects from the 4 different areas CDFS/UDS GROUND/HST (each sample comprising about 40 galaxies) and compared the results. The different corrections obtained are within 5\% at all wavelength. Since this error is below the calibration accuracy achievable for spectroscopy, we decided to apply a unique correction to all spectra independently on the area they come from. 

\begin{figure}
\resizebox{\hsize}{!}{\includegraphics[clip=true]{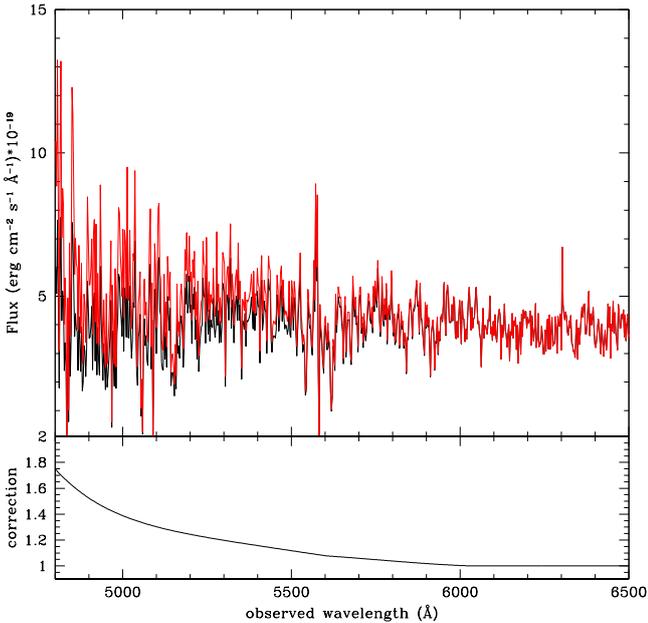}}
\caption{Top: uncorrected (black) and corrected (red) spectrum of galaxy CDFS114560. Bottom: the correction applied}
\label{bc}
\end{figure}

The top panel of Figure \ref{bc} shows the uncorrected data for objects CDFS114560 (randomly chosen from the spectroscopic catalogue) (black line)  and the spectrum after having applied the correction (red line), while the bottom panel shows the correction we have applied. Redwards of 6500\AA~ the correction is null, and its effects starts to be appreciable bluewards of 5500\AA.
In the distribution we include both the spectra corrected for the
blue flux loss, which we believe are our best calibration, and the spectra without the blue correction.  
\\
\begin{figure*}
\centering
\resizebox{\hsize}{!}{\includegraphics[bb=0 0 592 300,clip=true,width=17cm]{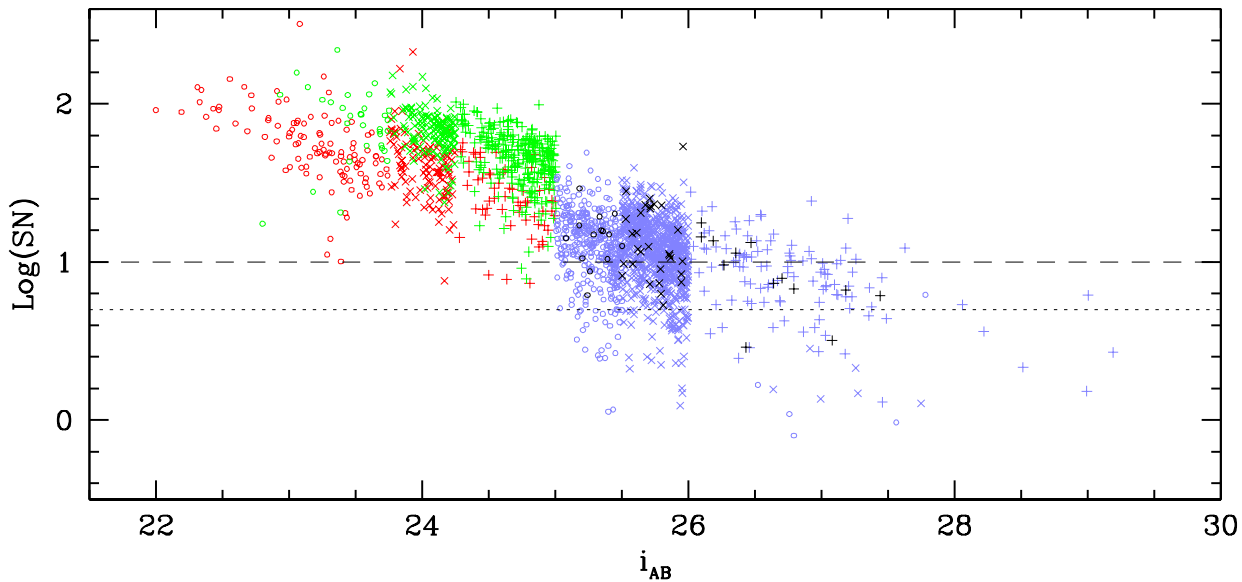} }
\caption{Signal to noise ratio per resolution element as a function of $i\mathrm{_{AB}}$ magnitude. Red symbols are the {\it passive} sample, green the SF sample, blue the {\it LBG} sample and black the {\it AGN} sample. Different symbol shapes indicate the different exposure times: circles for 20 hour, diagonal crosses for 40 hour and crosses for 80 hour exposures. The dotted line corresponds to S/N=5 per resolution element, while the dashed line to S/N=10}
\label{sn}
\end{figure*}
\\
To check the fullfillment of the requirements on S/N, we show in Figure~\ref{sn} the signal to noise ratio (S/N) per resolution element obtained on the final 1D spectra as a function of $i\mathrm{_{AB}}$ magnitude. Different symbols and colours are used for the different samples and exposure times: red for {\it passive} galaxies, green for {\it SF} galaxies, blue for {\it LBGs}, and black for {\it AGNs}, circles are for 20 hours, diagonal crosses for 40 hours and crosses for 80 hours exposure time. The S/N has been computed
as the mean S/N per resolution element
in the observed wavelength range 6500-7500~\AA~up to z=4. At higher redshifts we have used the observed redshift range 7500-8500~\AA~in order to remain redwards of the Lyman break.  Thanks to our nested observing strategy, the relation between log(S/N) and magnitude remains linear over the whole magnitude range (albeit with some scatter): basically all {\it passive} and {\it SF} galaxies (which have been selected to be  brigther than $i\mathrm{_{AB}=25}$) show a S/N higher than 10, while 85\% of the spectra of objects from the  {\it LBG} and {\it AGN} sample spectra show a S/N higher than the requsted value of S/N=5.
\\
\subsection{Redshift estimation, reliability flags and confidence levels}
\label{redshift}
 The redshift measurement strategy has been detailed in \citet{vandelsdr1}. In short, the redshift of each spectrum was measured using template fitting techniques or emission line measurements by 2 different team members, without knowledge of the photometric redshift. The two measures were reconciled and a provisional redshift flag assigned. 
 As a final step, all spectra were
independently re-checked by the two PIs and any remaining
discrepancies in the redshifts and quality flags were again reconciled.
This final step was necessary mainly to homogenize
 the quality flags as much as possible. The reliability of the measured redshifts is quantified following a scheme similar to that used for the VVDS \citep{vvds_main} and
zCosmos surveys \citep{zcosmos_main}.  Measurements of galaxies
are flagged using the following convention: 

\begin{itemize} 
\item
    Flag 4: a highly reliable redshift (estimated to have $\mathrm{>99\%}$
    probability of being correct), based on a high S/N
spectrum and supported by obvious and consistent spectral features.  
\item
    Flag 3: also a very reliable redshift, comparable in confidence
    with Flag 4, supported by clear spectral features in the spectrum, but not
necessarily with high S/N. 
\item
    Flag 2: a fairly reliable redshift measurement, but not as
      straightforward to confirm as for Flags 3 and 4,  supported by
    cross-correlation results, continuum shape and some spectral
    features, with an expected chance of $\mathrm{\simeq 75\%}$ of being
    correct. We shall see in the following that the actual estimated confidence
  level turns out to be significantly better.  
\item
    Flag 1: a reasonable redshift measurement, based on weak spectral features
    and/or continuum shape, for which there is roughly a $50\%$ chance that
    the redshift is actually wrong.
\item
    Flag 0: no reliable spectroscopic redshift measurement was possible.
\item
    Flag 9: a redshift based on only one single clear spectral emission
feature.
\item
  Flag -10: spectrum with clear problems in the observation or data
  processing phases. It can be a failure in the {\it vmmps} Sky to CCD
  conversion (especially at field corners), or a failed extraction,  or a bad sky subtraction because the object is too
  close to the edge of the slit. 
\end{itemize}
In section \ref{accuracy} we will countercheck  the reliability  of our flagging system.
A similar classification is used for broad line AGNs (BLAGNs). We define an object as BLAGN when one emission line is resolved at the spectral resolution of the survey, and they are easily identified during the redshift measurement process. 
 The flagging system for BLAGNs is similar, though not
identical, to the one adopted for stars and galaxies:
\begin{itemize} 
\item
    Flag 14: secure BLAGN with a $>99\%$ reliable redshift, including at least
    2 broad lines; 
\item
    Flag 13: secure BLAGN with good confidence redshift, based on
    one broad line and some faint additional feature; 
\item
    Flag 12: a $>75\%$ reliable redshift measurement, but lines are not
    significantly broad, might not be an BLAGN;
\item
    Flag 11: a tentative redshift measurement, with spectral features not
    significantly broad. 
\item
    Flag 19: secure BLAGN with one single reliable emission line feature,
    redshift based on this line only;
\end{itemize}

 At this stage, no attempt has been made to separate starburst galaxies from type 2, narrow line AGN.  The  complete catalogue of these sources, together with their characterization, will be published in Bongiorno et al. (in preparation). 

Serendipitous (also called secondary) objects appearing by chance within the slit of the main
target  are identified by adding a `2' in front of the main flag (thus a serendipitous galaxy spectrum with a highly reliable redshift will have flag 24, while a serendipitous BLAGN spectrum with a highly reliable redshift will have flag 214). 
\\
\section{The VANDELS final sample}
\label{spectroSample}
\begin{figure}
\resizebox{\hsize}{!}{\includegraphics[clip=true]{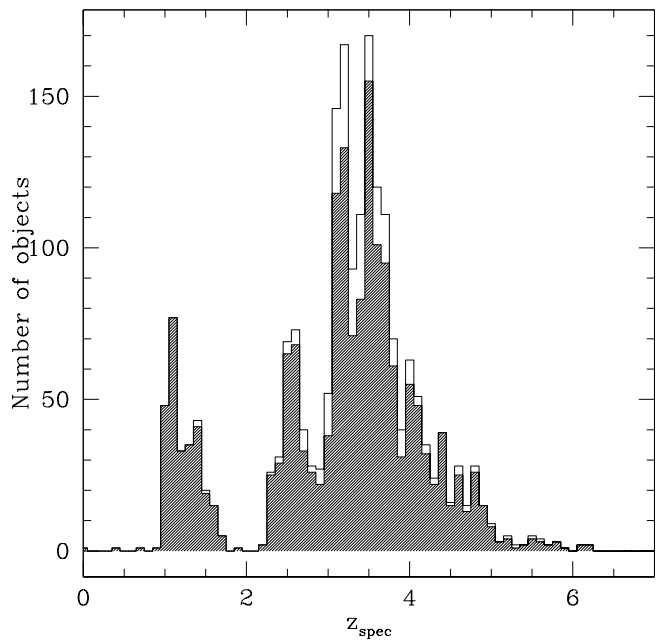} }
\caption{Redshift distribution of the final VANDELS sample: empty  histogram includes all measurements, shaded  one includes only secure redshifts.}
\label{zdist}
\end{figure}

Figure~\ref{zdist} shows the redshift distribution of the final VANDELS spectroscopic sample: shaded  for secure measurements (flags 2 through 9 and 12 through 19) while the empty histogram includes flags 1 and 11: the two distributions are very similar, showing that there has been no obvious redshift dependent bias in our redshift measurements. Table~\ref{surveyTable}, columns 5 and 6,  gives the number of measured redshifts and of secure measurements per object type and per area. Globally we have a redshift measurement for 2010 target galaxies, with a median redshift of $\mathrm{z=3.3}$. We underline that in Table~\ref{surveyTable} we refer to the {\it AGN} subsample as defined in section \ref{targetDef}. Almost all  the objects in the {\it AGN} subsample (i.e. those targets originally selected as potential AGNs) do not show Broad Lines in their spectra, and thus do not have a BLAGN spectroscopic redshift flag. Conversely, 16/17 objects with a BLAGN spectroscopic redshift flag do not belong to the original {\it AGN} subsample, but had been originally selected as {\it passive} or {\it SF} galaxies. A more detailed discussion on VANDELS AGNs and their spectroscopic properties will be presented in Bongiorno et al., in preparation.
\\
On top of the target sample, we have a non-negligible number of serendipitous objects for which a redshift could be measured. These are indicated as secondary objects in  Table \ref{surveyTable}. Overall, the VANDELS final data release comprises redshift and spectra for 2087 galaxies. As secondary objects usually do not fulfil our selection criteria, and the spectra are not of the same quality as the primary targets, we do not include them in the following analysis, but they are included in the release.

\begin{figure}
 \resizebox{\hsize}{!}{\includegraphics[clip=true]{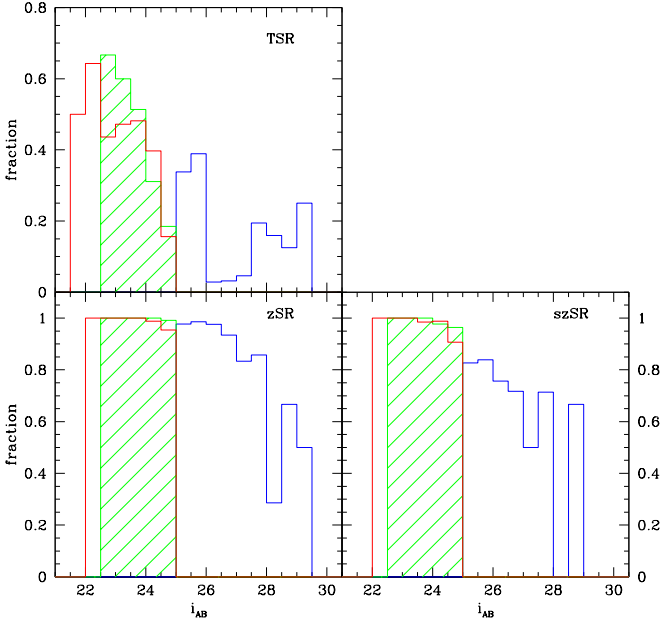}}
   \caption{Target Sampling Rate, TSR (top),  redshift measurement success rate, zSR (bottom left), and secure redshift measurement success rate, szSR, (bottom right)  for  {\it passive} (red), {\it SF} (green) and {\it LBG} samples (blue) as a function of $i\mathrm{_{AB}} $ magnitude.}
\label{tsr}
\end{figure}

VANDELS  has been conceived with the aim of providing a fair sample of high redshift galaxies, pre-selected on the basis of magnitude and photometric redshift. To assess how  representative of the parent photometric catalogue the spectroscopic sample is,
we define  Target Sampling Rate (TSR) as the fraction of observed galaxies over the parent sample, redshift  measurement Success Rate (zSR) as the fraction of galaxies with a measured redshift over the observed targets, and  secure redshift measurement Success Rate (szSR) as the fraction of objects with a secure redshift  over the observed targets. 
Looking at the total numbers in Table~\ref{surveyTable}, the TSR is roughly 20\% in both UDS and CDFS areas, but looking separately at the TSR of the three subsamples illustrated in Section \ref{design} as a function of $i\mathrm{_{AB}}$ magnitude (Figure~\ref{tsr}, top panel), we reach a  completeness larger than of 45\% for  {\it passive} galaxies down to  $i\mathrm{_{AB}}$=24.5 and for {\it SF} galaxies down to  $i\mathrm{_{AB}}$=24. The {\it LBG} sample  shows a TSR of $\mathrm{\sim}$ 40\%  down to  $i\mathrm{_{AB}}$=26, while for the faintest {\it LBGs}  the TSR drops  significantly, due to the 1:2:1 ratio in target selection we have applied during the mask preparation process described in Section \ref{maskdes}. On the other hand, given the total  exposure time allocated to the survey and the multiplexing capabilities of VIMOS when using  the MR grism (about 200 targets per pointing) on one side, and the long exposure times required for these faintest galaxies (80 hours), favouring a higher sampling rate would have drastically reduced the sampling of all the other kind of targets. 
\\
Looking at Table~\ref{surveyTable}, the global redshift measurement success rate zSR is 98\%, lowering to 86\% when only secure measurements are considered, with no difference between the UDS and CDFS fields.   Considering the three main samples of {\it passive}, {\it SF} and {\it LBGs}, Figure~\ref{tsr} (bottom left panel) shows zSR as a function of magnitude: it is higher than 95\% for all samples down to  $i\mathrm{_{AB}}$=27, and remains above 80\% even at $i\mathrm{_{AB}}$=28. Even limiting ourselves only to secure redshifts (Figure~\ref{tsr}, bottom right panel), the szSR  reaches almost 100\% for both {\it passive} and {\it SF} galaxies down to $i\mathrm{_{AB}}$=25, and remains above 70\% till $i\mathrm{_{AB}}$=27 for {\it LBGs}. This figure demonstrates the excellent quality of the VANDELS spectra even for the faintest and most distant galaxies we have targeted.
\\

 \subsection{Redshift accuracy, comparison with photometric redshifts and with literature data}
 \label{accuracy}
 Most of VANDELS targets have been observed for 40 hours or more, but many of them are detected with a decent S/N in  20 hour exposure time. Using DR1 and DR2   spectra (described in \citealt{vandelsdr1} and \citealt{vandelsdr2} respectively) we have 
 identified 283 objects observed for 20 hours in DR1 and 40 hours in DR2, and 193 objects observed for 40 hours in DR1 and 80 hours in DR2, plus two objects observed, respectively, for 40 and 80 hours in DR1 and 120 and 140 hours in DR2. Using the 478 double measures, we can assess whether the reliability level of our flagging system corresponds to what stated in section \ref{redshift}. The distribution of the differences between the redshifts independently measured from the spectra extracted from the data with two different exposure times  for all galaxies with spectroscopic flags = 2, 3, 4, and 9 is well represented by a Gaussian centred at 0 with $\mathrm{\sigma_{\Delta z/(1+z)} = 0.0007}$. From this sigma we estimate that the average redshift uncertainty of a single measurement is $\mathrm{\sigma/\sqrt {2}\sim147 km s^{-1}}$

 We define two redshift measurements as being in agreement when |$\Delta$z/(1+z) | < 0.020 (i.e. $\sim \mathrm{3\sigma}$ of the observed dispersion in the measurements). We indicate with $\mathrm{p_{i}}$, with i=1,2,3,4,9 the probability that the redshifts associated to each flag are correct (as from  Section \ref{redshift}), with $\mathrm{n_{tot_{i,j}}}$ the total number of pairs of measurements having  spectroscopic flags i and j, and with $\mathrm{n_{good_{i,j}}}$ the number of  pairs of measurements having  spectroscopic flags i and j which are in agreement with each other. 
 Applying the binomial distribution, the likelihood of getting the observed number of {\it good} redshifts in agreement in all the pairs with the various flags can be written as:

 \begin{equation}
\mathrm{L =  \Pi_{(i,j)}
B_{i,j}
\cdot (p_{i}p_{j})^{n_{good_{i,j}}} 
\cdot (1-p_{i}p_{j})^{n_{bad_{i,j}}} 
} 
\end{equation}
where 
 \begin{equation}
 \mathrm{
 B_{i,j}=\frac{n_{tot_{i,j}}!}{n_{good_{i,j}}! (n_{tot_{i,j}}-n_{good_{i,j}})!} 
 }
\end{equation}
and
$\mathrm{n_{bad_{i,j}}=n_{tot_{i,j}}-n_{good_{i,j}}}$
\\\\
Rearranging the factors and dropping the terms which do not depend on the reliabilities $\mathrm{p_{i}}$, the likelihood can be rewritten as:
\begin{equation}
\mathrm{L =  \Pi_{(i)}p_{i}^{expo_{i}} \cdot  \Pi_{(i,j)}(1-p_{i}p_{j})^{expo_{i,j}}}
\label{eq2}
\end{equation}
where \\
\\
$\mathrm{expo_{i} =2 \cdot n_{good_{i,i}} + \Sigma_{(j\neq i)}n_{good_{i,j}}}$ \\
$\mathrm{expo_{i,j} = n_{bad_{i,j}}}$ 
\\\\
The best estimates for the reliabilities $\mathrm{p_{i}}$ are computed by maximizing the likelihood in Equation \ref{eq2}, while their $\mathrm{1\sigma}$ uncertainties have been computed by projecting on each  $\mathrm{p_{i}}$ axis the surface with $\mathrm{\Delta S = S - S_{min} = 1}$, where S=-2lnL.
The results are shown in Table \ref{probTable}, where for each flag, we give the estimate of the  probability  for the redshift to be correct, as well as the 1 $\sigma$ range: we estimate a  reliability of  almost 100\% for flags 3 and 4, almost 80\% for flags 2, 95\% for flags 9. In Appendix \ref{appendix} we report the total number of double measurements  $\mathrm{n_{tot_{i,j}}}$ and  the number of good double measurements $\mathrm{n_{good_{i,j}}}$ for all flags. Flags 1 seem to have a reliability slightly lower than what assumed in Section \ref{redshift}. 
Overall, redshifts with the highest flags (3,4 and 9) have a confidence level above 95\%. 
\\
\begin{table}
\caption{Redshift flag measured reliability}
\label{probTable}
\centering
\begin{tabular}{c c c c}
\hline\hline
Flag       &  Measured reliability    &    1 $\sigma$ range\\
\hline
3-4         &         0.987           &       0.981$-$0.990 \\
2           &          0.79             &          0.75$-$0.83    \\            
1           &          0.41             &          0.36$-$0.45 \\
9           &          0.95             &          0.91$-$0.97  \\
\hline
 \end{tabular}
 \end{table}

\begin{figure}
\resizebox{\hsize}{!}{\includegraphics[clip=true]{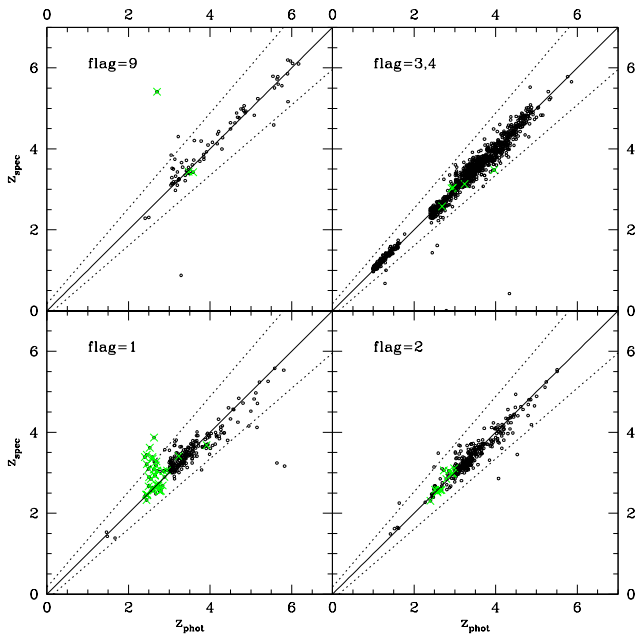} }
\caption{
Comparison between spectroscopic and photometric redshifts per reliability  flag. In each panel, black circles are for the {\it passive, SF} and {\it LBG} samples, superimposed green crosses mark the objects from the {\it AGN} sample. The solid line shows the 1:1 relation and the dotted lines mark the outlier limit ($abs (dz) >0.15$, where $\mathrm{dz=(z_{spec} - z_{phot})/(1 + z_{spec})}$).
}
\label{photoz}
\end{figure}

In Figure~\ref{photoz} we show the comparison between  spectroscopic 
redshifts and the photometric redshifts we had been using for the parent samples selection. Following \citet{mclure2018}, we define as bias the median value  of $\mathrm{dz=(z_{spec} - z_{phot})/(1 + z_{spec})}$ and as accuracy $\mathrm{\sigma(dz)}$ the Median Absolute Deviation (MAD) 
of the bias. Outliers are those objects showing $\mathrm{abs (dz) >0.15}$.  Considering the whole sample of measured redshifts, including all flags, the bias is 0.0011, with an accuracy of 0.019 and 1.5\% outliers. In Section \ref{targetDef} we have underlined how we expect the photometric redshifts of the {\it AGN} sample  to be less accurate than the bulk of the VANDELS targets, and this is confirmed by looking at Figure~\ref{photoz}: many objects from the {\it AGN} sample, represented as  superimposed green crosses, fall out of the  outlier limit. Indeed, excluding the {\it AGN} sample from the computation, the bias becomes smaller than $\mathrm{10^{-4}}$ and the accuracy lowers to 0.018. Outliers are 1\%. Considering only secure spectroscopic redshifts, these numbers do not change in a significant way.  This shows that the photometric redshifts used for our initial selection 
were robust, and their usage has not introduced unknown biases in the sample. 
\\

\begin{figure}
\resizebox{\hsize}{!}{\includegraphics[clip=true]{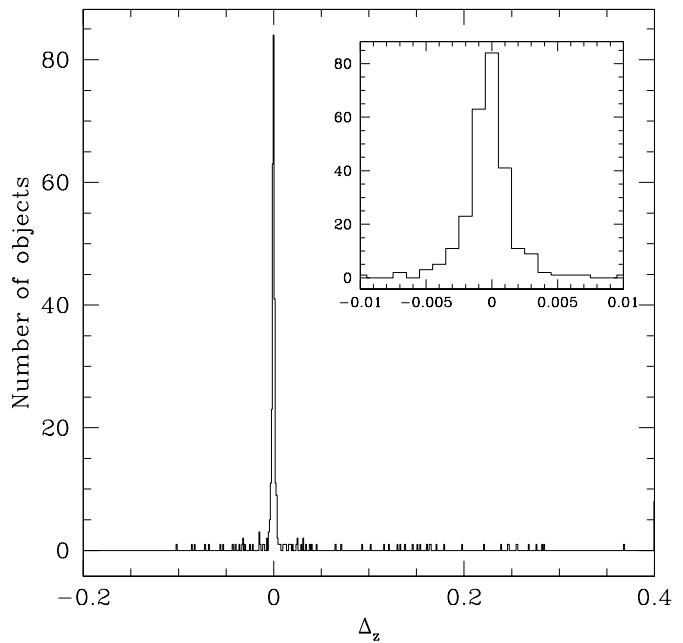} }
\caption{
Distribution of the differences  between VANDELS and previously published redshift values. The inset zooms into the central part of the histogram.
}
\label{compalitt}
\end{figure}

Among the VANDELS targets for which we have a redshift measurement, 336 objects have a redshift measurement already published in the literature. Comparison with literature values should be done using measurements obtained with similar wavelength resolution and of the same quality, which is not always straightforward as different authors use different quality estimators, as well as different measurement techniques: for example, in some surveys the redshift is based on emission line measurements, while we use also template fitting which accounts for both emission and absorption features, and this may introduce small differences.  
Nevertheless, we have compared all published values, irrespective of their quality, with our measurements, and the resulting distribution of redshift differences is shown in Figure \ref{compalitt}.
The distribution is very peaked (70\% of the measurements differ by less than 0.003) and well centreed on zero. Definining 
the bias and the accuracy in the same way as for the photometric redshifts, VANDELS measurements are in excellent  agreement with data from the literature, the bias being less than $\mathrm{10^{-4}}$  and the accuracy $\mathrm{\sim10^{-3}}$. 

\section{VANDELS spectra}
\begin{figure*}
\centering
\resizebox{\hsize}{!}{\includegraphics[clip=true,keepaspectratio ]{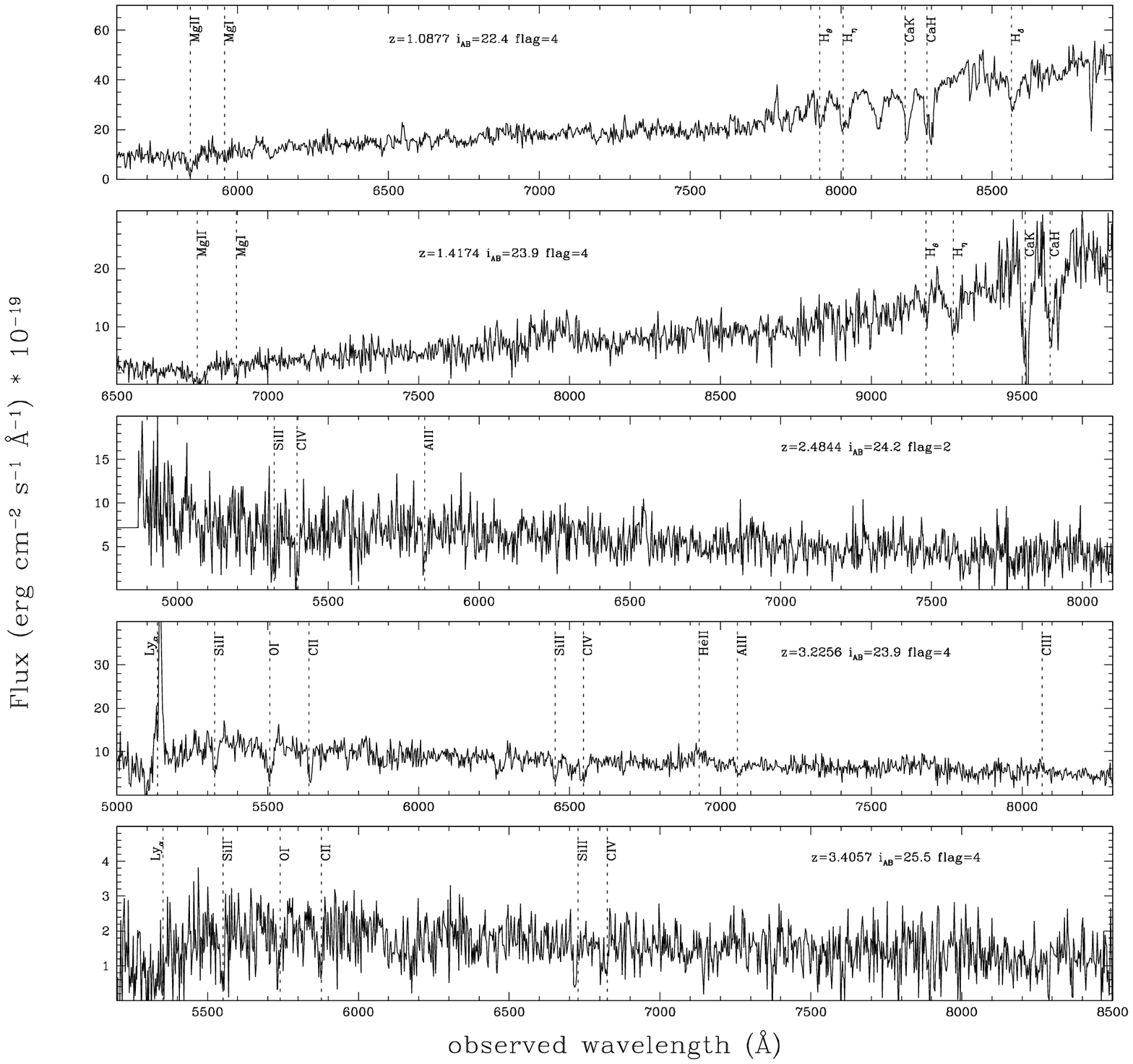} }
\caption{Examples of VANDELS spectra of galaxies drawn from different subsamples, magnitude and redshift ranges. Spectra are zoomed around the region containing the most prominent lines, according to the galaxy redshift. From top to bottom: two {\it passive} galaxies and three {\it SF} galaxies. The redshift, magnitude and reliability flag of each galaxy are indicated in each panel. }
\label{spectra1}
\end{figure*}

\begin{figure*}
\centering
\resizebox{\hsize}{!}{\includegraphics[bb=0 120 592 600, clip=true,keepaspectratio ]{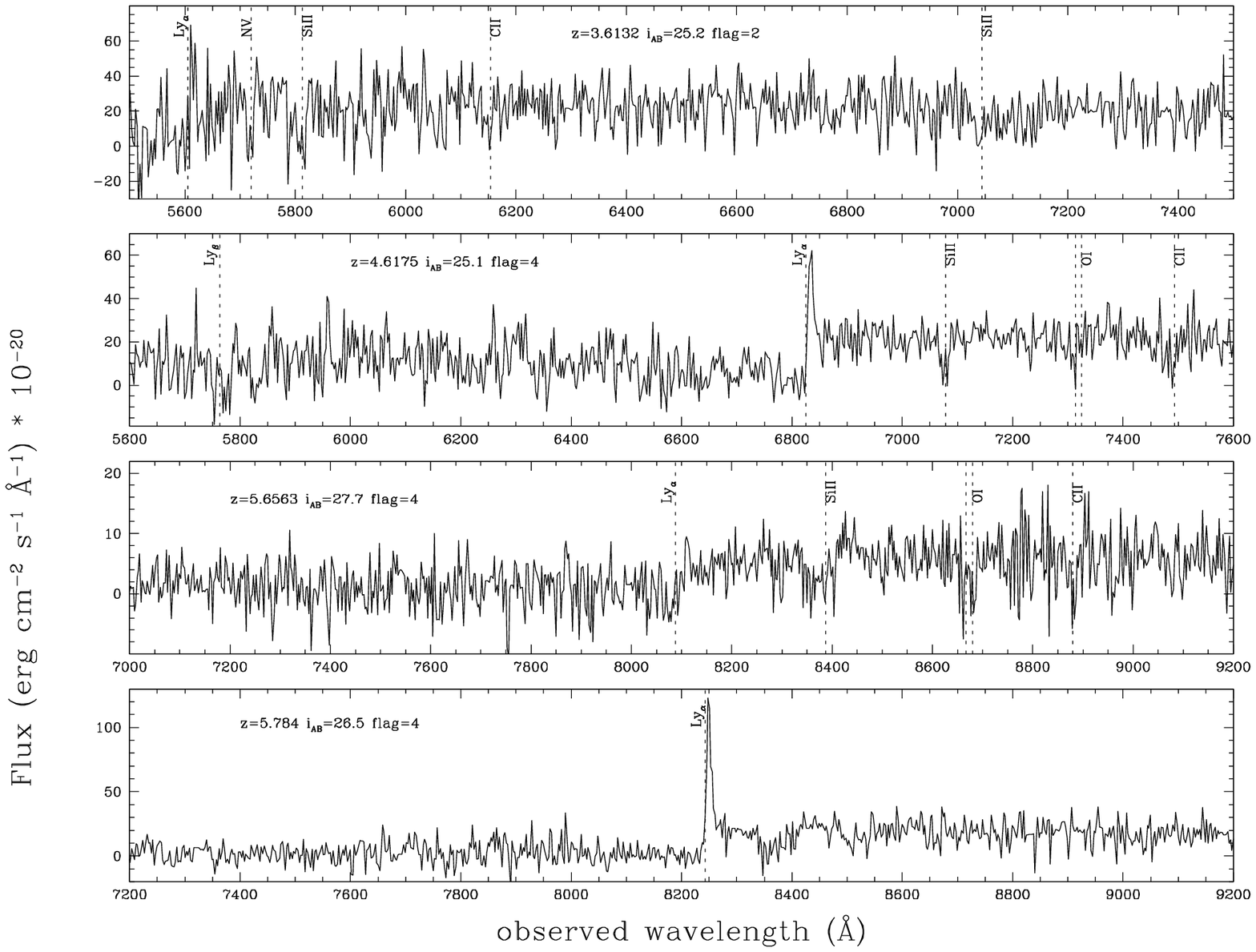} }
\caption{As Figure \ref{spectra1} for 4 {\it LBGs} in the redshift range 3.6 to 5.8.}
\label{spectra2}
\end{figure*}

Figures \ref{spectra1} and \ref{spectra2} show a few examples of VANDELS spectra of
galaxies from the different subsamples at different redshifts. 
To better show the quality of the data, we plot only the part of the spectrum with the stronger lines, according to the redshift of the galaxy.
All spectra have been normalized to the object $i\mathrm{_{AB}}$ magnitude and corrected for the blue drop (see Section \ref{reduction}). In Figure \ref{spectra1}, from top to bottom, we show spectra for {\it passive} galaxies at $\mathrm{z\sim1}$ and $\mathrm{z\sim1.4}$, and for {\it SF} galaxies at $\mathrm{z\sim2.4}$, 3.2 and 3.4. Figure \ref{spectra2} is dedicated to {\it LBGs}, from $\mathrm{z\sim3.6}$ to $\mathrm{z\sim5.8}$. Magnitudes range from the relatively bright values  of {\it passive} galaxies ($i_{AB}$ from 22.3 to 23.9) to the faint {\it LBGs}, the faintest object we show here has $i\mathrm{_{AB}\sim27.7}$.  The two figures  demonstrate the exquisite  quality of the spectra, and the wealth of information which can be derived from them even in the faintest and furthest away objects.

 \section{Intrinsic galaxy properties}
\begin{figure}
 \resizebox{\hsize}{!}{\includegraphics[clip=true]{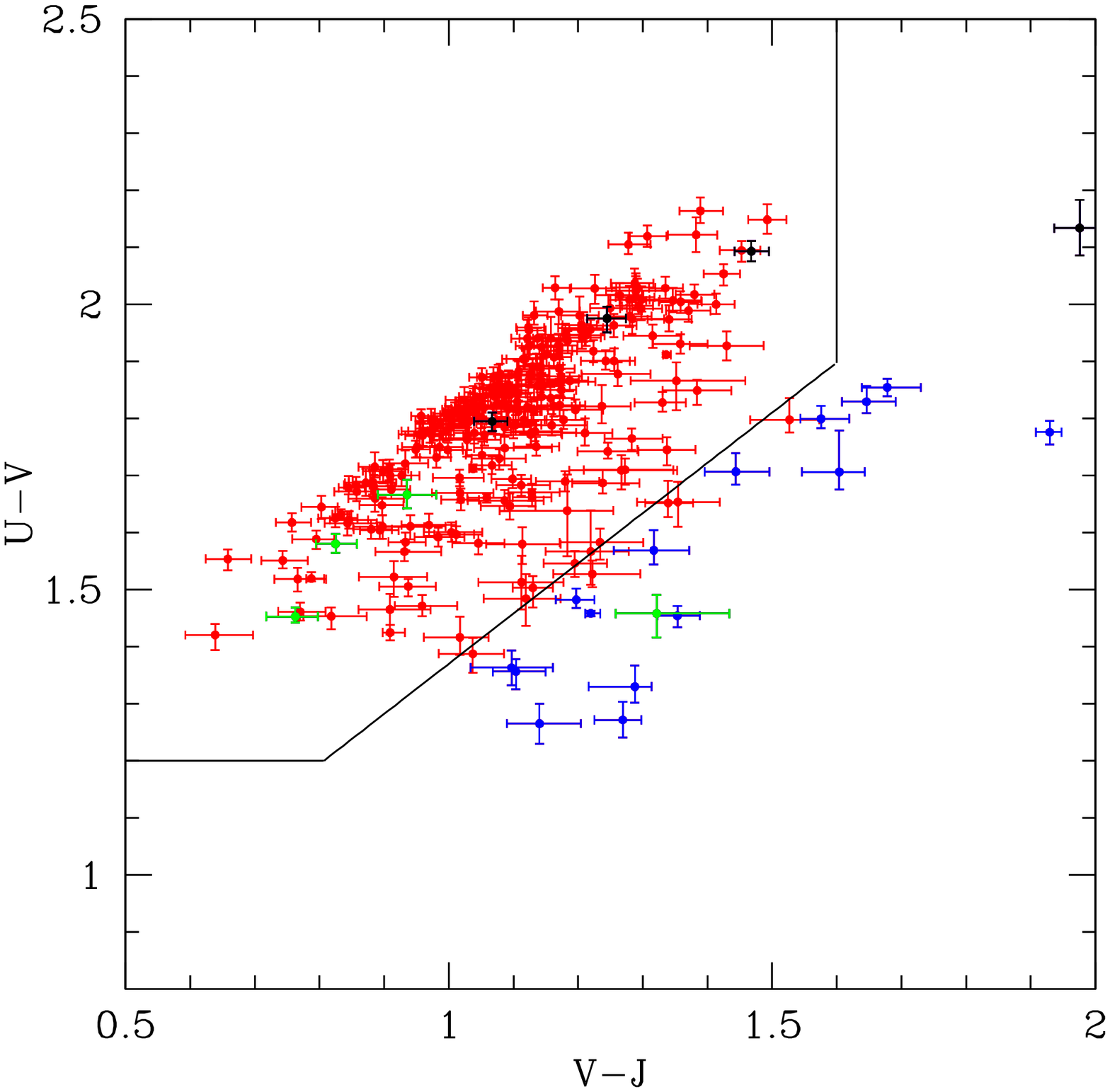}}
  \caption{UVJ diagram for the {\it passive} sample: the black lines indicate the {\it passive} galaxy selection box \citep{williams}. Black symbols for objects turned out to be interlopers, green for objects classified as BLAGNs, blue for previously classified passive objects now falling outside the selection box.}
\label{uvj}
\end{figure}

Coupling  the long baseline of the photometric coverage with the  excellent redshift measurement quality, we can perform SED fitting  to derive the physical properties and the corresponding uncertainties of our spectroscopic sample. At this stage, SED fitting has been performed using {\sc BAGPIPES} \citep{bagpipes}, fixing the redshift at the spectroscopically measured value and using all the new available ground-based photometry.
With respect to the catalogs described in Section \ref{parent}, the new ground based catalogues feature deeper near-IR data, fully deconfused Spitzer IRAC photometry and improved PSF homogenization.
\\
The {\sc BAGPIPES} code was run using a simplified configuration designed to mimic that used by \citet{mclure2018} when selecting the VANDELS sample. We use the 2016 updated version of the \citet{bc03} models using the MILES stellar spectral  library \citep{falconBarroso}  and updated stellar evolutionary tracks of \citet{bressan}  and \citet{marigo}. The Star Formation History (SFH) is parameterised using an exponentially declining form with a minimum timescale of 10 Myr and minimum age of 50 Myr. The stellar metallicity was fixed to the Solar value and no emission lines were included in the fitting process. Dust attenuation was modelled using the \citet{calzetti} model, with a maximum A$_{V}$ = 2.5 magnitudes. Whilst this model configuration is similar to what used by other large public surveys when publishing physical parameter catalogues, it should be noted that the details of the model used can have a substantial impact on the results obtained (e.g. \citealt{carnall2019}, \citealt{leja}).
\\
We check whether our original classification, made on a previous version of the photometric catalogues, using photometric redshifts as described in Section \ref{parent} and \ref{targetDef} and based on a different SED fitting code, still holds using spectroscopic redshifts, improved photometry and {\sc BAGPIPES} results. Figure \ref{uvj} shows the new UVJ diagram for the subsample of {\it passive} galaxies. 
Out of the 278 {\it passive} galaxies for which a redshift has been measured, 250 still satisfy the colour colour selection criterion (red dots), 4 turned out to be at a redshift below the selection range (green dots) (3 out of 4 are at redshift between 0.97 and 0.98), 4 have been spectroscopically classified as broad line AGNs (black dots), and 20 are not compatible any more with the selection locus (blue dots). We note that from the initial selection of the sample, in the SED fitting we now use the higher quality photometry available and the spectroscopic redshifts instead of the photometric ones used for pre-selection. Coupled with the usage of {\sc BAGPIPES} for SED fitting, this explains the 10\% change in classification we observe.
Galaxies formerly selected as {\it passive} and now falling outside the selection box had already been identified in \citet{carnall2019}, 
and classified as Post Starburst Galaxies on the basis of the SED fitting parameters. Indeed, these galaxies show a higher SFR than the {\it passive} ones, 80\% of them having log(SFR)>1 \msun/y. 
Among the 417 spectroscopically measured  {\it SF} galaxies, 5 objects fall out of the redshift selection criterion once the photometric redshift precision (0.15, see section \ref{accuracy}) is accounted for, while 4 turned out to be Broad Line AGNs.  Similarly among the 1259 {\it LBG} measured objects, 8 do not satify the redshift range criteria and 8 are Broad Line AGNs. The Specific SFR criterion remains satisfied for all other galaxies.

\begin{figure}
 \resizebox{\hsize}{!}{\includegraphics[clip=true]{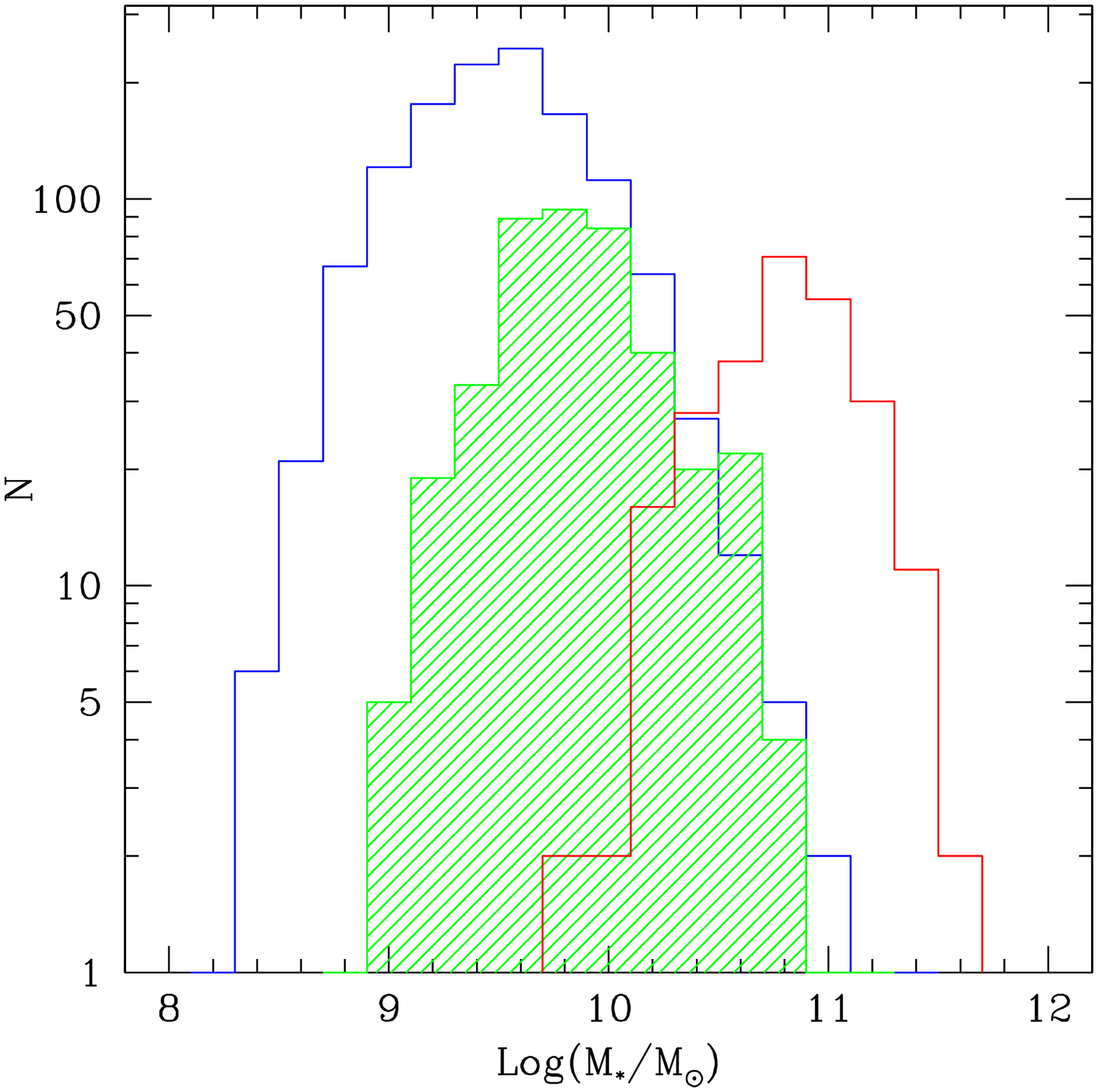}}
  \caption{Stellar mass distribution  for the three subsamples of {\it passive} (red), {\it SF} (green) and {\it LBG} (blue) galaxies.}
\label{mass}
\end{figure}

\begin{figure}
 \resizebox{\hsize}{!}{\includegraphics[clip=true]{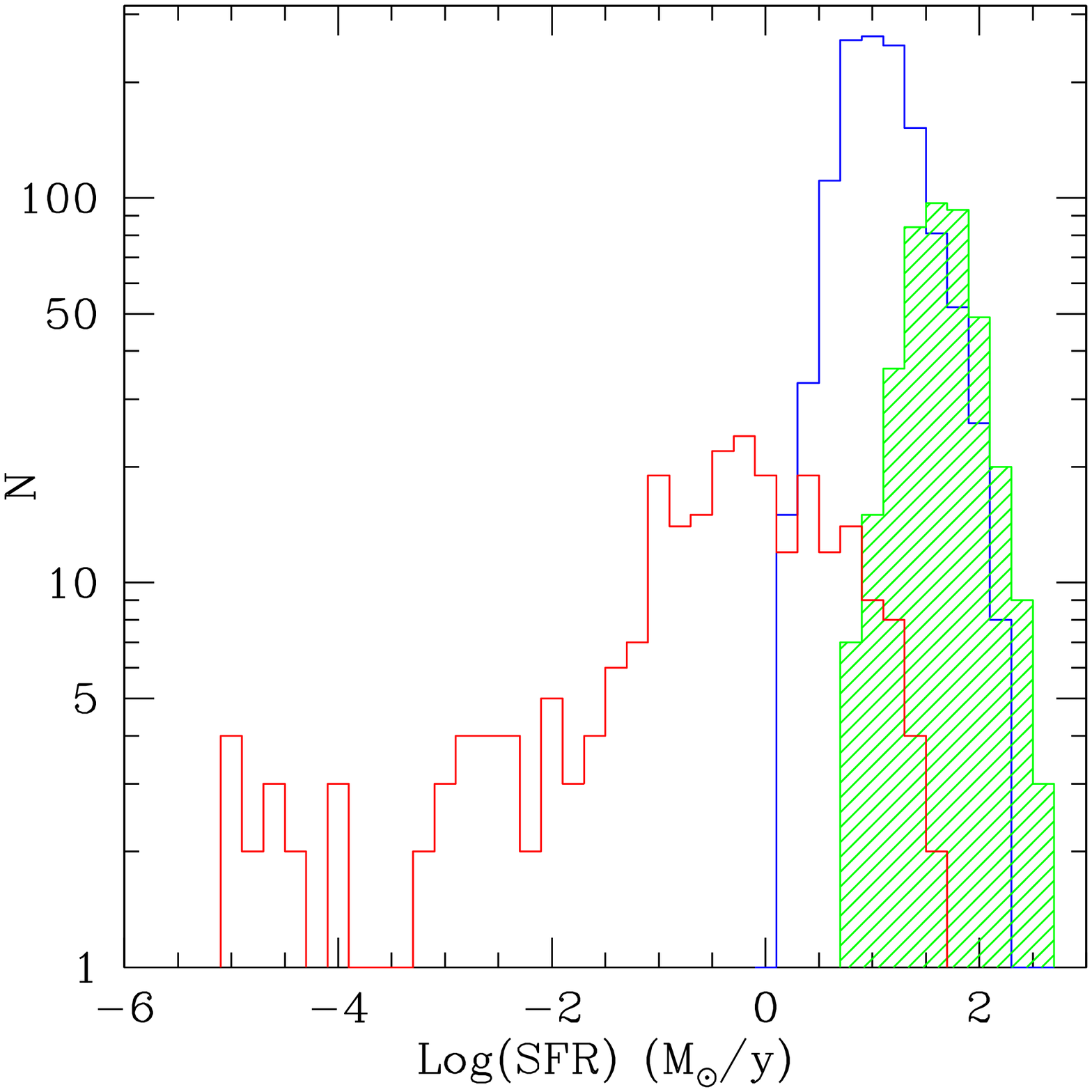}}
  \caption{SFR  distribution for the three subsamples of {\it passive} (red), {\it SF} (green) and {\it LBG} (blue) galaxies.}
\label{sfr}
\end{figure}

Figure~\ref{mass} shows the stellar mass distribution of the spectroscopic sample, divided in the three subsamples of {\it passive} (red), {\it SF (green)} and {\it LBG} (blue) galaxies as redefined with the new SED fitting. We span the mass range between Log($\mathrm{M_{*}}/$\msun)=8.3, and Log($\mathrm{M_{*}}/$\msun)=11.7, with the {\it passive} galaxies  dominating above Log($\mathrm{M_{*}}/$\msun)=10.8.
 Figure~\ref{sfr} shows the SFR distribution. The {\it passive} galaxies subsample as redefined on the basis of Figure \ref{uvj} dominates the low SFRs. Furthermore, 92\% of the 250 objects in the UVJ selection box also satisfy the sSFR<0.1 $\mathrm{Gyr^{-1} }$ condition and all of them have sSFR<0.5 $\mathrm{Gyr^{-1} }$.

\begin{figure*}
\center
\includegraphics[bb=0 0 592 250, width=17cm,clip=true]{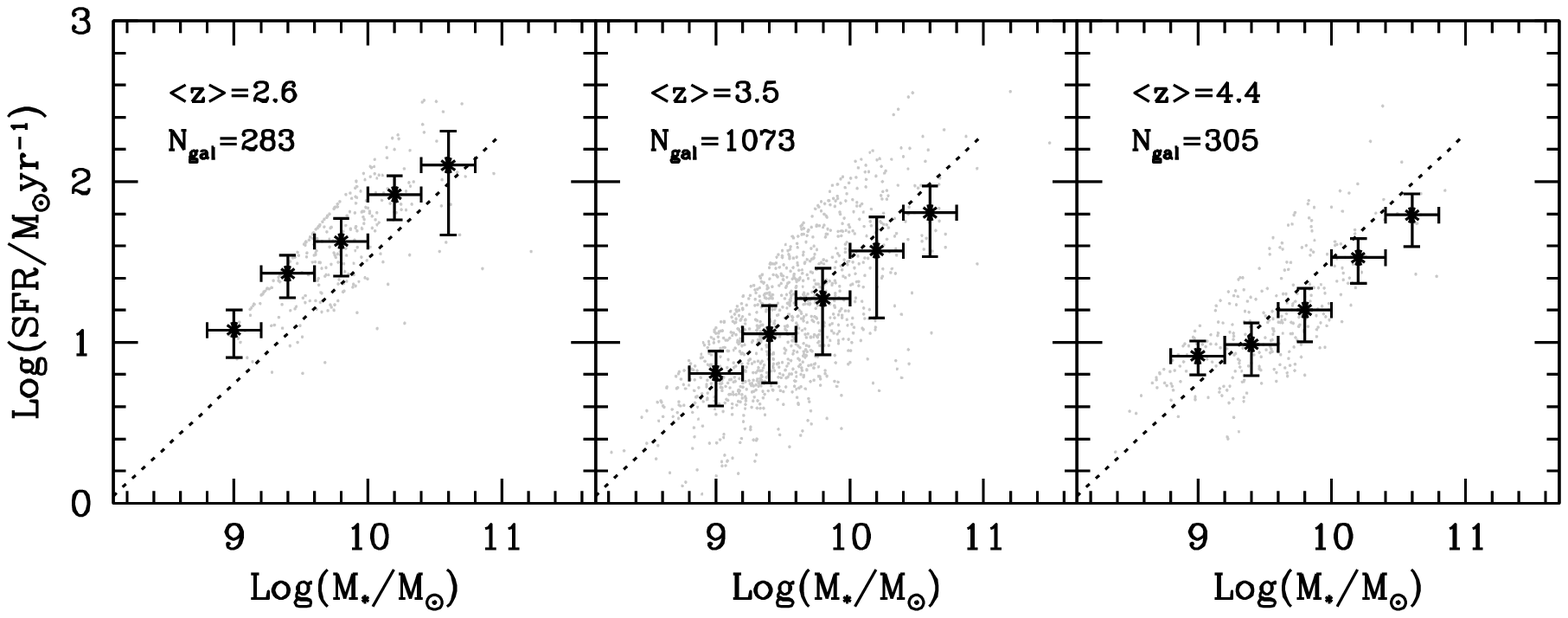} 
  \caption{The SFR-M${_*}$ plane for {\it SF} galaxies and {\it LBGs} in the VANDELS survey: left, z<3; centre, 3<z<4; right z>4. The median redshift  is indicated in the plot, as well as the number of galaxies in each redshift bin. Light dots are the individual galaxies,  big stars show the median SFR in mass bins. In each panel the dotted  line is  the relation by \citet{speagle}, computed at the median redshift of the bin.}
\label{ms}
\end{figure*}

The SFR-M${_*}$ plane for {\it SF} galaxies and {\it LBGs} is shown in Figure \ref{ms} for three different redshift ranges: 2<z<3 (left), 3<z<4 (middle), z>4 (right). We overplot the median SFR values (stars) computed in mass bins. Error bars on the x axis show the mass bins  width, while error bars on the y axis are the Median Absolute Deviation of the SFR within that mass bin.  The dotted line line is the relation by \citet{speagle} computed at the median redshift of the sample in each redshift range. In the lowest redshift range, <z>=2.6, VANDELS measurements are above the relation by \citet{speagle}. This is a result of our selection criterion: only bright (i.e. $i\mathrm{_{AB} \le 25}$) {\it SF} galaxies enter the redshift bin 2.4<z<3 (cf. section \ref{targetDef}). Given the magnitude limited selection, these galaxies are the brightest in the UV rest-frame and this explains why they are mainly above the Main Sequence. In the redshift range 3<z<4, where the observed sample is constituted by both bright {\it SF} galaxies and fainter {\it LBGs},  our values are in good agreement with the relation by \citet{speagle}, confirming the results of \citet{cullen2018}  obtained with the first VANDELS data release. In the highest redshift range the sample is dominated by faint {\it LBGs}, and our points are slightly below, but still compatible with the relation by \citet{speagle} at these redshifts.

\section{Public Data Release and Database Access}
\label{database}
The public data release comprises: 
\begin{enumerate}
\item Catalogues, for UDS and CDFS areas separately, containing spectroscopic results (Table~\ref{spectroCat}), photometric measurements (Table~\ref{photom}) and SED fitting results (Table~\ref{sedTable}); 
\item Spectra: the  reduced and calibrated 1D spectra and the resampled and sky subtracted (but not flux calibrated) 2D spectra.
\end{enumerate} 
Both catalogues and spectra  are available from the VANDELS consortium site ({\tt http://vandels.inaf.it/db}) as well as from the ESO catalog facility ({\tt https://www.eso.org/qi/}), the only difference being the spectral format. From  the VANDELS consortium site, 1D and 2D spectra for a single object can be downloaded  as a single multi-extension FITS file containing the following extensions:
\begin{itemize}
\item Primary: the 1D spectrum in $\mathrm{erg~cm^{-2} s^{-1} \AA^{-1}}$, with the blue end correction applied. 
\item EXR2D: the 2D linearly resampled spectrum in counts
\item SKY: the subtracted 1D sky spectrum in counts
\item NOISE: the 1D noise estimate in $\mathrm{erg~cm^{-2} s^{-1} \AA^{-1}}$.
\item EXR1D: a copy of the Primary 1D spectrum (to recover any editing that might be done on the Primary)
\item THUMB: the image thumbnail of the object
\item EXR1D\_UNCORR: the original 1D spectrum (see Section \ref{reduction})
\end{itemize}
where each 1D spectrum is a mono dimensional image (i.e the standard IRAF and/or IDL image format).\\
From the  ESO archive, 1D spectra can be downloaded as VO-table like FITS files, i.e. each spectrum is a FITS binary table containing the following columns: 
  \begin{itemize}
  \item WAVE: wavelength in Angstroms (in air)
  \item FLUX: 1D spectrum flux in $\mathrm{erg~cm^{-2} s^{-1} \AA^{-1}}$
  \item ERR: noise estimate in $\mathrm{erg~cm^{-2}s^{-1} \AA^{-1}}$
  \item UNCORR\_FLUX: 1D spectrum flux uncorrected for blue flux loss (see Section \ref{reduction})
  \item SKY: the subtracted sky in counts
 \end{itemize}
while the 2D spectra are distributed as separated FITS images.\\ 

\begin{table*}
\caption{catalogue contents }
\label{spectroCat}
\centering
\begin{tabular}{c c}
\hline\hline
Name  & Description   \\
\hline
id     &   Object identification        \\
alpha     & J2000 Right Ascension in decimal degrees      \\
delta     & J2000 Declination in decimal degrees      \\
zspec     & Spectroscopic redshift \\
zflg     &  redshift confidence flag as described in Section~\ref{redshift}        \\
Photometric catalogue & HST or GROUND \\
fluxes & optical and NIR fluxes in $\mathrm{\mu Jy}$ as described in Table \ref{photom} \\
Object properties & SED fitting results, see table \ref{sedTable}\\
\hline
\end{tabular}
\end{table*}

\begin{table*}
\caption{Distributed Photometry }
\label{photom}
\centering
\begin{tabular}{c c}
\hline\hline
CDFS  & UDS   \\
\hline\multicolumn{2}{c}{Ground based photometric catalog} \\
\hline
U (VIMOS) & U (CFHT) \\
B (WFI)  &  \\
IA484, IA527 IA598 IA624 IA651 IA679 IA738 IA767 (Subaru) & B,V,R,i z (Subaru) \\
F606W & NB921 (Subaru) \\
R (VIMOS) & \\
F850LP & \\
Z, Y, J, H, Ks (VISTA) &Y (VISTA), J,H,K (WFCAM) \\
CH1, CH2 (IRAC) & CH1, CH2 (IRAC) \\
\hline
\multicolumn{2}{c}{HST photometric catalog} \\
\hline
U (VIMOS) & U CFHT\\
 & B, V, R, i ,z (Subaru) \\
F435W, F606W F775W F814W F850LP F098M F105W F125W F160W & F606W, F125W, F160W \\
 & Y (HAWKI) \\
 & J ,H,K (WFCAM) \\
Ks (ISAAC) & \\
Ks (HAWKI) & Ks (HAWKI) \\
CH1, CH2 (IRAC) & CH1, CH2 (IRAC) \\
\hline
\end{tabular}
\end{table*}

\begin{table*}
\caption{SED derived parameters }
\label{sedTable}
\centering
\begin{tabular}{c c}
\hline\hline
Name  & Description   \\
\hline
Av & V-band dust attenuation in magnitudes \\
age &  time since the onset of star formation in Gyr \\
massformed  & total stellar mass formed by the time of observation \\
tau & exponential timescale for the SFH in Gyr \\
stellar mass  & mass in living stars and remnants at the time of observation  \\
sfr & SFR averaged over the last 100 Myr  \\
ssfr  & SFR divided by stellar mass   \\
UV colour & rest-frame U-V colour using the filter curves described in \citet{williams} \\
VJ colour & rest-frame V-J colour using the filter curves described in \citet{williams} \\
chisq phot & raw minimum chi-squared value for the fit to the data \\
n bands & number of photometric bands used in the fit \\
\hline
\end{tabular}
\end{table*}

\section{Summary}
We present the final Public Data  Release of the VANDELS survey, which includes  2087  redshifts of
galaxies in the range $\mathrm{1<z<6.5}$. Complementing the general description
given in \citet{mclure2018} and in \citet{vandelsdr1},  
we discuss the details of the target selection, observations, data reduction and
redshift measurements, providing all relevant information for the proper use of the
data.  

Thanks to the extremely deep observations (up to 80 hours of exposure time), the signal to noise per resolution element of the spectra is above 7 for 80\% of the targets with a magnitude brighter than $i\mathrm{_{AB}=26}$, while 70\% of the spectra of fainter targets have a signal to noise higher than 5. 

The VANDELS survey spans the redshift range 1<z<6.5, with a target sampling rate greater than 45\% for {\it passive} galaxies down to $\mathrm{i_{AB}=24.5}$ and for {\it SF} galaxies down to $\mathrm{i_{AB}=24}$. The spectroscopic measurement success rate is as high as 98\% considering all redshift measurements, and 86\% considering only redshifts with a reliability above 80\%. By internal comparison between different observations, we estimate a redshift accuracy of 0.0007.

We have performed  SED fitting to derive galaxy intrinsic properties. The sample covers the mass range 8.3<Log(M$_{*}$/M$_\odot$)<11.7. We show that neither the target selection process or the redshift measurement process has introduced further significant biases with respect to the original selection based on photometric redshifts.

The full spectroscopic catalogues, together with the complementary
photometric information and SED fitting derived  quantities are publicly
available from  {\tt  http://vandels.inaf.it}, as well as from the ESO archive {\tt https://www.eso.org/qi/}. 
Measurements of line fluxes, equivalent widths and Lick indexes will be made available in the near future.

\begin{acknowledgements}
We thank an anonymous referee for the useful comments which helped improving the quality of the paper.
This work has been partially funded by Premiale MITIC 2017 and INAF PRIN "Mainstream 2019". AC acknowledges the support from grant PRIN MIUR 2017-20173ML3WW-001; RA acknowledges support from ANID FONDECYT Regular 1202007. FB acknowledges Junta de Castilla y Le\'{o}n and the European Regional Development Fund (ERDF) for financial support under grant BU229P18. MM and AC acknowledge support from the grants ASI n.I/023/12/0, ASI n.2018- 23-HH.0. MM acknowledges support from MIUR, PRIN 2017 (grant 20179ZF5KS). MF acknowledges support from the UK Science and Technology Facilities Council (STFC) (grant number ST/R000905/1).
The TOPCAT software \citep{topcat} has been widely use for this paper.
\end{acknowledgements}

\bibliographystyle{aa}
\bibliography{garilli_vandels}

\begin{thebibliography}{72}
\expandafter\ifx\csname natexlab\endcsname\relax\def\natexlab#1{#1}\fi

\bibitem[{{Abazajian} {et~al.}(2003){Abazajian}, {Adelman-McCarthy},
  {Ag{\"u}eros}, {Allam}, {Anderson}, {Annis}, {Bahcall}, {Baldry}, {Bastian},
  {Berlind}, {Bernardi}, {Blanton}, {Blythe}, {Bochanski}, {Boroski},
  {Brewington}, {Briggs}, {Brinkmann}, {Brunner}, {Budav{\'a}ri}, {Carey},
  {Carr}, {Castander}, {Chiu}, {Collinge}, {Connolly}, {Covey}, {Csabai},
  {Dalcanton}, {Dodelson}, {Doi}, {Dong}, {Eisenstein}, {Evans}, {Fan},
  {Feldman}, {Finkbeiner}, {Friedman}, {Frieman}, {Fukugita}, {Gal},
  {Gillespie}, {Glazebrook}, {Gonzalez}, {Gray}, {Grebel}, {Grodnicki}, {Gunn},
  {Gurbani}, {Hall}, {Hao}, {Harbeck}, {Harris}, {Harris}, {Harvanek},
  {Hawley}, {Heckman}, {Helmboldt}, {Hendry}, {Hennessy}, {Hindsley}, {Hogg},
  {Holmgren}, {Holtzman}, {Homer}, {Hui}, {Ichikawa}, {Ichikawa}, {Inkmann},
  {Ivezi{\'c}}, {Jester}, {Johnston}, {Jordan}, {Jordan}, {Jorgensen},
  {Juri{\'c}}, {Kauffmann}, {Kent}, {Kleinman}, {Knapp}, {Kniazev}, {Kron},
  {Krzesi{\'n}ski}, {Kunszt}, {Kuropatkin}, {Lamb}, {Lampeitl}, {Laubscher},
  {Lee}, {Leger}, {Li}, {Lidz}, {Lin}, {Loh}, {Long}, {Loveday}, {Lupton},
  {Malik}, {Margon}, {McGehee}, {McKay}, {Meiksin}, {Miknaitis}, {Moorthy},
  {Munn}, {Murphy}, {Nakajima}, {Narayanan}, {Nash}, {Neilsen}, {Newberg},
  {Newman}, {Nichol}, {Nicinski}, {Nieto-Santisteban}, {Nitta}, {Odenkirchen},
  {Okamura}, {Ostriker}, {Owen}, {Padmanabhan}, {Peoples}, {Pier}, {Pindor},
  {Pope}, {Quinn}, {Rafikov}, {Raymond}, {Richards}, {Richmond}, {Rix},
  {Rockosi}, {Schaye}, {Schlegel}, {Schneider}, {Schroeder}, {Scranton},
  {Sekiguchi}, {Seljak}, {Sergey}, {Sesar}, {Sheldon}, {Shimasaku}, {Siegmund},
  {Silvestri}, {Sinisgalli}, {Sirko}, {Smith}, {Smol{\v c}i{\'c}}, {Snedden},
  {Stebbins}, {Steinhardt}, {Stinson}, {Stoughton}, {Strateva}, {Strauss},
  {SubbaRao}, {Szalay}, {Szapudi}, {Szkody}, {Tasca}, {Tegmark}, {Thakar},
  {Tremonti}, {Tucker}, {Uomoto}, {Vanden Berk}, {Vandenberg}, {Vogeley},
  {Voges}, {Vogt}, {Walkowicz}, {Weinberg}, {West}, {White}, {Wilhite},
  {Willman}, {Xu}, {Yanny}, {Yarger}, {Yasuda}, {Yip}, {Yocum}, {York},
  {Zakamska}, {Zehavi}, {Zheng}, {Zibetti}, \& {Zucker}}]{sdss1}
{Abazajian}, K., {Adelman-McCarthy}, J.~K., {Ag{\"u}eros}, M.~A., {et~al.}
  2003, \aj, 126, 2081

\bibitem[{{Ahumada} {et~al.}(2020){Ahumada}, {Allende Prieto}, {Almeida},
  {Anders}, {Anderson}, {Andrews}, {Anguiano}, {Arcodia}, {Armengaud},
  {Aubert}, {Avila}, {Avila-Reese}, {Badenes}, {Balland }, {Barger},
  {Barrera-Ballesteros}, {Basu}, {Bautista}, {Beaton}, {Beers}, {Benavides},
  {Bender}, {Bernardi}, {Bershady}, {Beutler}, {Bidin}, {Bird}, {Bizyaev},
  {Blanc}, {Blanton}, {Boquien}, {Borissova}, {Bovy}, {Brand t}, {Brinkmann},
  {Brownstein}, {Bundy}, {Bureau}, {Burgasser}, {Burtin}, {Cano-D{\'\i}az},
  {Capasso}, {Cappellari}, {Carrera}, {Chabanier}, {Chaplin}, {Chapman},
  {Cherinka}, {Chiappini}, {Doohyun Choi}, {Chojnowski}, {Chung}, {Clerc},
  {Coffey}, {Comerford}, {Comparat}, {da Costa}, {Cousinou}, {Covey}, {Crane},
  {Cunha}, {da Silva Ilha}, {Dai}, {Damsted}, {Darling}, {Davidson}, {Davies},
  {Dawson}, {De}, {de la Macorra}, {De Lee}, {de Andrade Queiroz}, {Deconto
  Machado}, {de la Torre}, {Dell'Agli}, {du Mas des Bourboux},
  {Diamond-Stanic}, {Dillon}, {Donor}, {Drory}, {Duckworth}, {Dwelly},
  {Ebelke}, {Eftekharzadeh}, {Eigenbrot}, {Elsworth}, {Eracleous},
  {Erfanianfar}, {Escoffier}, {Fan}, {Farr}, {Fern{\'a}ndez-Trincado},
  {Feuillet}, {Finoguenov}, {Fofie}, {Fraser-McKelvie}, {Frinchaboy},
  {Fromenteau}, {Fu}, {Galbany}, {Garcia}, {Garc{\'\i}a-Hern{\'a}ndez}, {Garma
  Oehmichen}, {Ge}, {Geimba Maia}, {Geisler}, {Gelfand }, {Goddy},
  {Gonzalez-Perez}, {Grabowski}, {Green}, {Grier}, {Guo}, {Guy}, {Harding},
  {Hasselquist}, {Hawken}, {Hayes}, {Hearty}, {Hekker}, {Hogg}, {Holtzman},
  {Horta}, {Hou}, {Hsieh}, {Huber}, {Hunt}, {Ider Chitham}, {Imig}, {Jaber},
  {Jimenez Angel}, {Johnson}, {Jones}, {J{\"o}nsson}, {Jullo}, {Kim},
  {Kinemuchi}, {Kirkpatrick}, {Kite}, {Klaene}, {Kneib}, {Kollmeier}, {Kong},
  {Kounkel}, {Krishnarao}, {Lacerna}, {Lan}, {Lane}, {Law}, {Le Goff}, {Leung},
  {Lewis}, {Li}, {Lian}, {Lin}, {Long}, {Longa-Pe{\~n}a}, {Lundgren}, {Lyke},
  {Ted Mackereth}, {MacLeod}, {Majewski}, {Manchado}, {Maraston}, {Martini},
  {Masseron}, {Masters}, {Mathur}, {McDermid}, {Merloni}, {Merrifield},
  {M{\'e}sz{\'a}ros}, {Miglio}, {Minniti}, {Minsley}, {Miyaji}, {Mohammad},
  {Mosser}, {Mueller}, {Muna}, {Mu{\~n}oz-Guti{\'e}rrez}, {Myers}, {Nadathur},
  {Nair}, {Nandra}, {do Nascimento}, {Nevin}, {Newman}, {Nidever}, {Nitschelm},
  {Noterdaeme}, {O'Connell}, {Olmstead}, {Oravetz}, {Oravetz}, {Osorio},
  {Pace}, {Padilla}, {Palanque-Delabrouille}, {Palicio}, {Pan}, {Pan},
  {Parker}, {Paviot}, {Peirani}, {Pe{\~n}a Ram{\'r}ez}, {Penny}, {Percival},
  {Perez-Fournon}, {P{\'e}rez-R{\`a}fols}, {Petitjean}, {Pieri},
  {Pinsonneault}, {Poovelil}, {Povick}, {Prakash}, {Price-Whelan}, {Raddick},
  {Raichoor}, {Ray}, {Rembold}, {Rezaie}, {Riffel}, {Riffel}, {Rix}, {Robin},
  {Roman-Lopes}, {Rom{\'a}n-Z{\'u}{\~n}iga}, {Rose}, {Ross}, {Rossi}, {Rowland
  s}, {Rubin}, {Salvato}, {S{\'a}nchez}, {S{\'a}nchez-Menguiano},
  {S{\'a}nchez-Gallego}, {Sayres}, {Schaefer}, {Schiavon}, {Schimoia},
  {Schlafly}, {Schlegel}, {Schneider}, {Schultheis}, {Schwope}, {Seo},
  {Serenelli}, {Shafieloo}, {Shamsi}, {Shao}, {Shen}, {Shetrone}, {Shirley},
  {Silva Aguirre}, {Simon}, {Skrutskie}, {Slosar}, {Smethurst}, {Sobeck},
  {Sodi}, {Souto}, {Stark}, {Stassun}, {Steinmetz}, {Stello}, {Stermer},
  {Storchi-Bergmann}, {Streblyanska}, {Stringfellow}, {Stutz}, {Su{\'a}rez},
  {Sun}, {Taghizadeh-Popp}, {Talbot}, {Tayar}, {Thakar}, {Theriault}, {Thomas},
  {Thomas}, {Tinker}, {Tojeiro}, {Toledo}, {Tremonti}, {Troup}, {Tuttle},
  {Unda-Sanzana}, {Valentini}, {Vargas-Gonz{\'a}lez}, {Vargas-Maga{\~n}a},
  {V{\'a}zquez-Mata}, {Vivek}, {Wake}, {Wang}, {Weaver}, {Weijmans}, {Wild},
  {Wilson}, {Wilson}, {Wolthuis}, {Wood-Vasey}, {Yan}, {Yang}, {Y{\`e}che},
  {Zamora}, {Zarrouk}, {Zasowski}, {Zhang}, {Zhao}, {Zhao}, {Zheng}, {Zheng},
  {Zhu}, \& {Zou}}]{sdss16}
{Ahumada}, R., {Allende Prieto}, C., {Almeida}, A., {et~al.} 2020, \apjs, 249,
  3

\bibitem[{{Bacon} {et~al.}(2017){Bacon}, {Conseil}, {Mary}, {Brinchmann},
  {Shepherd}, {Akhlaghi}, {Weilbacher}, {Piqueras}, {Wisotzki}, {Lagattuta},
  {Epinat}, {Guerou}, {Inami}, {Cantalupo}, {Courbot}, {Contini}, {Richard},
  {Maseda}, {Bouwens}, {Bouch{\'e}}, {Kollatschny}, {Schaye}, {Marino},
  {Pello}, {Herenz}, {Guiderdoni}, \& {Carollo}}]{muse}
{Bacon}, R., {Conseil}, S., {Mary}, D., {et~al.} 2017, \aap, 608, A1

\bibitem[{{Bertin} \& {Arnouts}(1996)}]{sextractor}
{Bertin}, E. \& {Arnouts}, S. 1996, \aaps, 117, 393

\bibitem[{{Bielby} {et~al.}(2011){Bielby}, {Shanks}, {Weilbacher}, {Infante},
  {Crighton}, {Bornancini}, {Bouch{\'e}}, {H{\'e}raudeau}, {Lambas},
  {Lowenthal}, {Minniti}, {Padilla}, {Petitjean}, \& {Theuns}}]{VLT_LBG}
{Bielby}, R.~M., {Shanks}, T., {Weilbacher}, P.~M., {et~al.} 2011, \mnras, 414,
  2

\bibitem[{{Bottini} {et~al.}(2005){Bottini}, {Garilli}, {Maccagni}, {Tresse},
  {Le Brun}, {Le F{\`e}vre}, {Picat}, {Scaramella}, {Scodeggio}, {Vettolani},
  {Zanichelli}, {Adami}, {Arnaboldi}, {Arnouts}, {Bardelli}, {Bolzonella},
  {Cappi}, {Charlot}, {Ciliegi}, {Contini}, {Foucaud}, {Franzetti}, {Guzzo},
  {Ilbert}, {Iovino}, {McCracken}, {Marano}, {Marinoni}, {Mathez}, {Mazure},
  {Meneux}, {Merighi}, {Paltani}, {Pollo}, {Pozzetti}, {Radovich}, {Zamorani},
  \& {Zucca}}]{vmmps}
{Bottini}, D., {Garilli}, B., {Maccagni}, D., {et~al.} 2005, \pasp, 117, 996

\bibitem[{{Brammer} {et~al.}(2012){Brammer}, {van Dokkum}, {Franx},
  {Fumagalli}, {Patel}, {Rix}, {Skelton}, {Kriek}, {Nelson}, {Schmidt},
  {Bezanson}, {da Cunha}, {Erb}, {Fan}, {F{\"o}rster Schreiber}, {Illingworth},
  {Labb{\'e}}, {Leja}, {Lundgren}, {Magee}, {Marchesini}, {McCarthy},
  {Momcheva}, {Muzzin}, {Quadri}, {Steidel}, {Tal}, {Wake}, {Whitaker}, \&
  {Williams}}]{brammer2012}
{Brammer}, G.~B., {van Dokkum}, P.~G., {Franx}, M., {et~al.} 2012, \apjs, 200,
  13

\bibitem[{{Bressan} {et~al.}(2012){Bressan}, {Marigo}, {Girardi}, {Salasnich},
  {Dal Cero}, {Rubele}, \& {Nanni}}]{bressan}
{Bressan}, A., {Marigo}, P., {Girardi}, L., {et~al.} 2012, \mnras, 427, 127

\bibitem[{{Bruzual} \& {Charlot}(2003)}]{bc03}
{Bruzual}, G. \& {Charlot}, S. 2003, \mnras, 344, 1000

\bibitem[{{Burgasser}(2014)}]{Burgasser}
{Burgasser}, A.~J. 2014, in Astronomical Society of India Conference Series,
  Vol.~11, Astronomical Society of India Conference Series, 7--16

\bibitem[{{Calabr{\`o}} {et~al.}(2020){Calabr{\`o}}, {Castellano},
  {Pentericci}, {Fontanot}, {Menci}, {Cullen}, {McLure}, {Bolzonella},
  {Cimatti}, {Marchi}, {Talia}, {Amor{\'\i}n}, {Cresci}, {De Lucia}, {Fynbo},
  {Fontana}, {Franco}, {Hathi}, {Hibon}, {Hirschmann}, {Mannucci}, {Santini},
  {Saxena}, {Schaerer}, {Xie}, \& {Zamorani}}]{calabro}
{Calabr{\`o}}, A., {Castellano}, M., {Pentericci}, L., {et~al.} 2020, arXiv
  e-prints, arXiv:2011.06615

\bibitem[{{Calzetti} {et~al.}(2000){Calzetti}, {Armus}, {Bohlin}, {Kinney},
  {Koornneef}, \& {Storchi-Bergmann}}]{calzetti}
{Calzetti}, D., {Armus}, L., {Bohlin}, R.~C., {et~al.} 2000, \apj, 533, 682

\bibitem[{{Carnall} {et~al.}(2019){Carnall}, {McLure}, {Dunlop}, {Cullen},
  {McLeod}, {Wild}, {Johnson}, {Appleby}, {Dav{\'e}}, {Amorin}, {Bolzonella},
  {Castellano}, {Cimatti}, {Cucciati}, {Gargiulo}, {Garilli}, {Marchi},
  {Pentericci}, {Pozzetti}, {Schreiber}, {Talia}, \& {Zamorani}}]{carnall2019}
{Carnall}, A.~C., {McLure}, R.~J., {Dunlop}, J.~S., {et~al.} 2019, \mnras, 490,
  417

\bibitem[{{Carnall} {et~al.}(2018){Carnall}, {McLure}, {Dunlop}, \&
  {Dav{\'e}}}]{bagpipes}
{Carnall}, A.~C., {McLure}, R.~J., {Dunlop}, J.~S., \& {Dav{\'e}}, R. 2018,
  \mnras, 480, 4379

\bibitem[{{Carnall} {et~al.}(2020){Carnall}, {Walker}, {McLure}, {Dunlop},
  {McLeod}, {Cullen}, {Wild}, {Amorin}, {Bolzonella}, {Castellano}, {Cimatti},
  {Cucciati}, {Fontana}, {Gargiulo}, {Garilli}, {Jarvis}, {Pentericci},
  {Pozzetti}, {Zamorani}, {Calabro}, {Hathi}, \& {Koekemoer}}]{carnall2020}
{Carnall}, A.~C., {Walker}, S., {McLure}, R.~J., {et~al.} 2020, \mnras, 496,
  695

\bibitem[{{Chabrier}(2003)}]{chabrier}
{Chabrier}, G. 2003, \pasp, 115, 763

\bibitem[{{Chang} {et~al.}(2017){Chang}, {Le Floc'h}, {Juneau}, {da Cunha},
  {Salvato}, {Civano}, {Marchesi}, {Ilbert}, {Toba}, {Lim}, {Tang}, {Wang},
  {Ferraro}, {Urry}, {Griffiths}, \& {Kartaltepe}}]{chang}
{Chang}, Y.-Y., {Le Floc'h}, E., {Juneau}, S., {et~al.} 2017, \apjs, 233, 19

\bibitem[{{Cimatti} {et~al.}(2002){Cimatti}, {Mignoli}, {Daddi}, {Pozzetti},
  {Fontana}, {Saracco}, {Poli}, {Renzini}, {Zamorani}, {Broadhurst},
  {Cristiani}, {D'Odorico}, {Giallongo}, {Gilmozzi}, \& {Menci}}]{k20}
{Cimatti}, A., {Mignoli}, M., {Daddi}, E., {et~al.} 2002, \aap, 392, 395

\bibitem[{{Colless} {et~al.}(2001){Colless}, {Dalton}, {Maddox}, {Sutherland},
  {Norberg}, {Cole}, {Bland-Hawthorn}, {Bridges}, {Cannon}, {Collins}, {Couch},
  {Cross}, {Deeley}, {De Propris}, {Driver}, {Efstathiou}, {Ellis}, {Frenk},
  {Glazebrook}, {Jackson}, {Lahav}, {Lewis}, {Lumsden}, {Madgwick}, {Peacock},
  {Peterson}, {Price}, {Seaborne}, \& {Taylor}}]{2df}
{Colless}, M., {Dalton}, G., {Maddox}, S., {et~al.} 2001, \mnras, 328, 1039

\bibitem[{{Cullen} {et~al.}(2020){Cullen}, {McLure}, {Dunlop}, {Carnall},
  {McLeod}, {Shapley}, {Amor{\'\i}n}, {Bolzonella}, {Castellano}, {Cimatti},
  {Cirasuolo}, {Cucciati}, {Fontana}, {Fontanot}, {Garilli}, {Guaita},
  {Jarvis}, {Pentericci}, {Pozzetti}, {Talia}, {Zamorani}, {Calabr{\`o}},
  {Cresci}, {Fynbo}, {Hathi}, {Giavalisco}, {Koekemoer}, {Mannucci}, \&
  {Saxena}}]{cullen2020}
{Cullen}, F., {McLure}, R.~J., {Dunlop}, J.~S., {et~al.} 2020, \mnras, 495,
  1501

\bibitem[{{Cullen} {et~al.}(2019){Cullen}, {McLure}, {Dunlop}, {Khochfar},
  {Dav{\'e}}, {Amor{\'\i}n}, {Bolzonella}, {Carnall}, {Castellano}, {Cimatti},
  {Cirasuolo}, {Cresci}, {Fynbo}, {Fontanot}, {Gargiulo}, {Garilli}, {Guaita},
  {Hathi}, {Hibon}, {Mannucci}, {Marchi}, {McLeod}, {Pentericci}, {Pozzetti},
  {Shapley}, {Talia}, \& {Zamorani}}]{cullen2019}
{Cullen}, F., {McLure}, R.~J., {Dunlop}, J.~S., {et~al.} 2019, \mnras, 487,
  2038

\bibitem[{{Cullen} {et~al.}(2018){Cullen}, {McLure}, {Khochfar}, {Dunlop},
  {Dalla Vecchia}, {Carnall}, {Bourne}, {Castellano}, {Cimatti}, {Cirasuolo},
  {Elbaz}, {Fynbo}, {Garilli}, {Koekemoer}, {Marchi}, {Pentericci}, {Talia}, \&
  {Zamorani}}]{cullen2018}
{Cullen}, F., {McLure}, R.~J., {Khochfar}, S., {et~al.} 2018, \mnras, 476, 3218

\bibitem[{{Daddi} {et~al.}(2004){Daddi}, {Cimatti}, {Renzini}, {Fontana},
  {Mignoli}, {Pozzetti}, {Tozzi}, \& {Zamorani}}]{Daddi04}
{Daddi}, E., {Cimatti}, A., {Renzini}, A., {et~al.} 2004, \apj, 617, 746

\bibitem[{{Falc{\'o}n-Barroso} {et~al.}(2011){Falc{\'o}n-Barroso},
  {S{\'a}nchez-Bl{\'a}zquez}, {Vazdekis}, {Ricciardelli}, {Cardiel}, {Cenarro},
  {Gorgas}, \& {Peletier}}]{falconBarroso}
{Falc{\'o}n-Barroso}, J., {S{\'a}nchez-Bl{\'a}zquez}, P., {Vazdekis}, A.,
  {et~al.} 2011, \aap, 532, A95

\bibitem[{{Galametz} {et~al.}(2013){Galametz}, {Grazian}, {Fontana},
  {Ferguson}, {Ashby}, {Barro}, {Castellano}, {Dahlen}, {Donley}, {Faber},
  {Grogin}, {Guo}, {Huang}, {Kocevski}, {Koekemoer}, {Lee}, {McGrath}, {Peth},
  {Willner}, {Almaini}, {Cooper}, {Cooray}, {Conselice}, {Dickinson}, {Dunlop},
  {Fazio}, {Foucaud}, {Gardner}, {Giavalisco}, {Hathi}, {Hartley}, {Koo},
  {Lai}, {de Mello}, {McLure}, {Lucas}, {Paris}, {Pentericci}, {Santini},
  {Simpson}, {Sommariva}, {Targett}, {Weiner}, {Wuyts}, \& {CANDELS
  Team}}]{Galametz}
{Galametz}, A., {Grazian}, A., {Fontana}, A., {et~al.} 2013, \apjs, 206, 10

\bibitem[{{Garilli} {et~al.}(2014){Garilli}, {Guzzo}, {Scodeggio},
  {Bolzonella}, {Abbas}, {Adami}, {Arnouts}, {Bel}, {Bottini}, {Branchini},
  {Cappi}, {Coupon}, {Cucciati}, {Davidzon}, {De Lucia}, {de la Torre},
  {Franzetti}, {Fritz}, {Fumana}, {Granett}, {Ilbert}, {Iovino}, {Krywult}, {Le
  Brun}, {Le F{\`e}vre}, {Maccagni}, {Ma{\l}ek}, {Marulli}, {McCracken},
  {Paioro}, {Polletta}, {Pollo}, {Schlagenhaufer}, {Tasca}, {Tojeiro},
  {Vergani}, {Zamorani}, {Zanichelli}, {Burden}, {Di Porto}, {Marchetti},
  {Marinoni}, {Mellier}, {Moscardini}, {Nichol}, {Peacock}, {Percival},
  {Phleps}, \& {Wolk}}]{vipers_dr1}
{Garilli}, B., {Guzzo}, L., {Scodeggio}, M., {et~al.} 2014, \aap, 562, A23

\bibitem[{{Garilli} {et~al.}(2008){Garilli}, {Le F{\`e}vre}, {Guzzo},
  {Maccagni}, {Le Brun}, {de la Torre}, {Meneux}, {Tresse}, {Franzetti},
  {Zamorani}, {Zanichelli}, {Gregorini}, {Vergani}, {Bottini}, {Scaramella},
  {Scodeggio}, {Vettolani}, {Adami}, {Arnouts}, {Bardelli}, {Bolzonella},
  {Cappi}, {Charlot}, {Ciliegi}, {Contini}, {Foucaud}, {Gavignaud}, {Ilbert},
  {Iovino}, {Lamareille}, {McCracken}, {Marano}, {Marinoni}, {Mazure},
  {Merighi}, {Paltani}, {Pell{\`o}}, {Pollo}, {Pozzetti}, {Radovich}, {Zucca},
  {Blaizot}, {Bongiorno}, {Cucciati}, {Mellier}, {Moreau}, \&
  {Paioro}}]{vvds_wide}
{Garilli}, B., {Le F{\`e}vre}, O., {Guzzo}, L., {et~al.} 2008, \aap, 486, 683

\bibitem[{{Garilli} {et~al.}(2012){Garilli}, {Paioro}, {Scodeggio},
  {Franzetti}, {Fumana}, \& {Guzzo}}]{easylife}
{Garilli}, B., {Paioro}, L., {Scodeggio}, M., {et~al.} 2012, \pasp, 124, 1232

\bibitem[{{Grogin} {et~al.}(2011){Grogin}, {Kocevski}, {Faber}, {Ferguson},
  {Koekemoer}, {Riess}, {Acquaviva}, {Alexander}, {Almaini}, {Ashby}, {Barden},
  {Bell}, {Bournaud}, {Brown}, {Caputi}, {Casertano}, {Cassata}, {Castellano},
  {Challis}, {Chary}, {Cheung}, {Cirasuolo}, {Conselice}, {Roshan Cooray},
  {Croton}, {Daddi}, {Dahlen}, {Dav{\'e}}, {de Mello}, {Dekel}, {Dickinson},
  {Dolch}, {Donley}, {Dunlop}, {Dutton}, {Elbaz}, {Fazio}, {Filippenko},
  {Finkelstein}, {Fontana}, {Gardner}, {Garnavich}, {Gawiser}, {Giavalisco},
  {Grazian}, {Guo}, {Hathi}, {H{\"a}ussler}, {Hopkins}, {Huang}, {Huang},
  {Jha}, {Kartaltepe}, {Kirshner}, {Koo}, {Lai}, {Lee}, {Li}, {Lotz}, {Lucas},
  {Madau}, {McCarthy}, {McGrath}, {McIntosh}, {McLure}, {Mobasher},
  {Moustakas}, {Mozena}, {Nandra}, {Newman}, {Niemi}, {Noeske}, {Papovich},
  {Pentericci}, {Pope}, {Primack}, {Rajan}, {Ravindranath}, {Reddy}, {Renzini},
  {Rix}, {Robaina}, {Rodney}, {Rosario}, {Rosati}, {Salimbeni}, {Scarlata},
  {Siana}, {Simard}, {Smidt}, {Somerville}, {Spinrad}, {Straughn}, {Strolger},
  {Telford}, {Teplitz}, {Trump}, {van der Wel}, {Villforth}, {Wechsler},
  {Weiner}, {Wiklind}, {Wild}, {Wilson}, {Wuyts}, {Yan}, \& {Yun}}]{candels2}
{Grogin}, N.~A., {Kocevski}, D.~D., {Faber}, S.~M., {et~al.} 2011, \apjs, 197,
  35

\bibitem[{{Guaita} {et~al.}(2020){Guaita}, {Pompei}, {Castellano},
  {Pentericci}, {Cucciati}, {Zamorani}, {Zoldan}, {Fontanot}, {Bauer},
  {Amorin}, {Bolzonella}, {de Lucia}, {Gargiulo}, {Hathi}, {Hibon},
  {Hirschmann}, {Koekemoer}, {McLure}, {Pozzetti}, {Talia}, {Thomas}, \&
  {Xie}}]{guaita}
{Guaita}, L., {Pompei}, E., {Castellano}, M., {et~al.} 2020, \aap, 640, A107

\bibitem[{{Guo} {et~al.}(2013){Guo}, {Ferguson}, {Giavalisco}, {Barro},
  {Willner}, {Ashby}, {Dahlen}, {Donley}, {Faber}, {Fontana}, {Galametz},
  {Grazian}, {Huang}, {Kocevski}, {Koekemoer}, {Koo}, {McGrath}, {Peth},
  {Salvato}, {Wuyts}, {Castellano}, {Cooray}, {Dickinson}, {Dunlop}, {Fazio},
  {Gardner}, {Gawiser}, {Grogin}, {Hathi}, {Hsu}, {Lee}, {Lucas}, {Mobasher},
  {Nand ra}, {Newman}, \& {van der Wel}}]{Guo}
{Guo}, Y., {Ferguson}, H.~C., {Giavalisco}, M., {et~al.} 2013, \apjs, 207, 24

\bibitem[{{Guzzo} {et~al.}(2014){Guzzo}, {Scodeggio}, {Garilli}, {Granett},
  {Fritz}, {Abbas}, {Adami}, {Arnouts}, {Bel}, {Bolzonella}, {Bottini},
  {Branchini}, {Cappi}, {Coupon}, {Cucciati}, {Davidzon}, {De Lucia}, {de la
  Torre}, {Franzetti}, {Fumana}, {Hudelot}, {Ilbert}, {Iovino}, {Krywult}, {Le
  Brun}, {Le F{\`e}vre}, {Maccagni}, {Ma{\l}ek}, {Marulli}, {McCracken},
  {Paioro}, {Peacock}, {Polletta}, {Pollo}, {Schlagenhaufer}, {Tasca},
  {Tojeiro}, {Vergani}, {Zamorani}, {Zanichelli}, {Burden}, {Di Porto},
  {Marchetti}, {Marinoni}, {Mellier}, {Moscardini}, {Nichol}, {Percival},
  {Phleps}, \& {Wolk}}]{vipers_main}
{Guzzo}, L., {Scodeggio}, M., {Garilli}, B., {et~al.} 2014, \aap, 566, A108

\bibitem[{{Hoag} {et~al.}(2019){Hoag}, {Treu}, {Pentericci}, {Amorin},
  {Bolzonella}, {Brada{\v{c}}}, {Castellano}, {Cullen}, {Fynbo}, {Garilli},
  {Guaita}, {Hathi}, {Henry}, {Jones}, {Mason}, {McLeod}, {McLure},
  {Morishita}, {Pozzetti}, {Schaerer}, {Schmidt}, {Talia}, \&
  {Thomas}}]{hoag2019}
{Hoag}, A., {Treu}, T., {Pentericci}, L., {et~al.} 2019, \mnras, 488, 706

\bibitem[{{Horne}(1986)}]{horne}
{Horne}, K. 1986, \pasp, 98, 609

\bibitem[{{Hsu} {et~al.}(2014){Hsu}, {Salvato}, {Nandra}, {Brusa}, {Bender},
  {Buchner}, {Donley}, {Kocevski}, {Guo}, {Hathi}, {Rangel}, {Willner},
  {Brightman}, {Georgakakis}, {Budav{\'a}ri}, {Szalay}, {Ashby}, {Barro},
  {Dahlen}, {Faber}, {Ferguson}, {Galametz}, {Grazian}, {Grogin}, {Huang},
  {Koekemoer}, {Lucas}, {McGrath}, {Mobasher}, {Peth}, {Rosario}, \&
  {Trump}}]{hsu}
{Hsu}, L.-T., {Salvato}, M., {Nandra}, K., {et~al.} 2014, \apj, 796, 60

\bibitem[{{Jarvis} {et~al.}(2013){Jarvis}, {Bonfield}, {Bruce}, {Geach},
  {McAlpine}, {McLure}, {Gonz{\'a}lez-Solares}, {Irwin}, {Lewis}, {Yoldas},
  {Andreon}, {Cross}, {Emerson}, {Dalton}, {Dunlop}, {Hodgkin}, {Le},
  {Karouzos}, {Meisenheimer}, {Oliver}, {Rawlings}, {Simpson}, {Smail},
  {Smith}, {Sullivan}, {Sutherland}, {White}, \& {Zwart}}]{video}
{Jarvis}, M.~J., {Bonfield}, D.~G., {Bruce}, V.~A., {et~al.} 2013, \mnras, 428,
  1281

\bibitem[{{Koekemoer} {et~al.}(2011){Koekemoer}, {Faber}, {Ferguson}, {Grogin},
  {Kocevski}, {Koo}, {Lai}, {Lotz}, {Lucas}, {McGrath}, {Ogaz}, {Rajan},
  {Riess}, {Rodney}, {Strolger}, {Casertano}, {Castellano}, {Dahlen},
  {Dickinson}, {Dolch}, {Fontana}, {Giavalisco}, {Grazian}, {Guo}, {Hathi},
  {Huang}, {van der Wel}, {Yan}, {Acquaviva}, {Alexander}, {Almaini}, {Ashby},
  {Barden}, {Bell}, {Bournaud}, {Brown}, {Caputi}, {Cassata}, {Challis},
  {Chary}, {Cheung}, {Cirasuolo}, {Conselice}, {Roshan Cooray}, {Croton},
  {Daddi}, {Dav{\'e}}, {de Mello}, {de Ravel}, {Dekel}, {Donley}, {Dunlop},
  {Dutton}, {Elbaz}, {Fazio}, {Filippenko}, {Finkelstein}, {Frazer}, {Gardner},
  {Garnavich}, {Gawiser}, {Gruetzbauch}, {Hartley}, {H{\"a}ussler},
  {Herrington}, {Hopkins}, {Huang}, {Jha}, {Johnson}, {Kartaltepe},
  {Khostovan}, {Kirshner}, {Lani}, {Lee}, {Li}, {Madau}, {McCarthy},
  {McIntosh}, {McLure}, {McPartland}, {Mobasher}, {Moreira}, {Mortlock},
  {Moustakas}, {Mozena}, {Nandra}, {Newman}, {Nielsen}, {Niemi}, {Noeske},
  {Papovich}, {Pentericci}, {Pope}, {Primack}, {Ravindranath}, {Reddy},
  {Renzini}, {Rix}, {Robaina}, {Rosario}, {Rosati}, {Salimbeni}, {Scarlata},
  {Siana}, {Simard}, {Smidt}, {Snyder}, {Somerville}, {Spinrad}, {Straughn},
  {Telford}, {Teplitz}, {Trump}, {Vargas}, {Villforth}, {Wagner}, {Wand ro},
  {Wechsler}, {Weiner}, {Wiklind}, {Wild}, {Wilson}, {Wuyts}, \&
  {Yun}}]{candels1}
{Koekemoer}, A.~M., {Faber}, S.~M., {Ferguson}, H.~C., {et~al.} 2011, \apjs,
  197, 36

\bibitem[{{Koo}(1995)}]{deep}
{Koo}, D. 1995, in Wide Field Spectroscopy and the Distant Universe, ed. S.~J.
  {Maddox} \& A.~{Aragon-Salamanca}, 55--+

\bibitem[{{Kurk} {et~al.}(2013){Kurk}, {Cimatti}, {Daddi}, {Mignoli},
  {Pozzetti}, {Dickinson}, {Bolzonella}, {Zamorani}, {Cassata}, {Rodighiero},
  {Franceschini}, {Renzini}, {Rosati}, {Halliday}, \& {Berta}}]{gmass}
{Kurk}, J., {Cimatti}, A., {Daddi}, E., {et~al.} 2013, \aap, 549, A63

\bibitem[{{Le F{\`e}vre} {et~al.}(2013){Le F{\`e}vre}, {Cassata}, {Cucciati},
  {Garilli}, {Ilbert}, {Le Brun}, {Maccagni}, {Moreau}, {Scodeggio}, {Tresse},
  {Zamorani}, {Adami}, {Arnouts}, {Bardelli}, {Bolzonella}, {Bondi},
  {Bongiorno}, {Bottini}, {Cappi}, {Charlot}, {Ciliegi}, {Contini}, {de la
  Torre}, {Foucaud}, {Franzetti}, {Gavignaud}, {Guzzo}, {Iovino}, {Lemaux},
  {L{\'o}pez-Sanjuan}, {McCracken}, {Marano}, {Marinoni}, {Mazure}, {Mellier},
  {Merighi}, {Merluzzi}, {Paltani}, {Pell{\`o}}, {Pollo}, {Pozzetti},
  {Scaramella}, {Tasca}, {Vergani}, {Vettolani}, {Zanichelli}, \&
  {Zucca}}]{vvds}
{Le F{\`e}vre}, O., {Cassata}, P., {Cucciati}, O., {et~al.} 2013, \aap, 559,
  A14

\bibitem[{{Le F{\`e}vre} {et~al.}(2003){Le F{\`e}vre}, {Saisse}, {Mancini},
  {Brau-Nogue}, {Caputi}, {Castinel}, {D'Odorico}, {Garilli}, {Kissler-Patig},
  {Lucuix}, {Mancini}, {Pauget}, {Sciarretta}, {Scodeggio}, {Tresse}, \&
  {Vettolani}}]{vvds_tech}
{Le F{\`e}vre}, O., {Saisse}, M., {Mancini}, D., {et~al.} 2003, in Proceedings
  of the SPIE, ed. M.~{Iye} \& A.~F.~M. {Moorwood}, Vol. 4841, 1670--1681

\bibitem[{{Le F{\`e}vre} {et~al.}(2015){Le F{\`e}vre}, {Tasca}, {Cassata},
  {Garilli}, {Le Brun}, {Maccagni}, {Pentericci}, {Thomas}, {Vanzella},
  {Zamorani}, {Zucca}, {Amorin}, {Bardelli}, {Capak}, {Cassar{\`a}},
  {Castellano}, {Cimatti}, {Cuby}, {Cucciati}, {de la Torre}, {Durkalec},
  {Fontana}, {Giavalisco}, {Grazian}, {Hathi}, {Ilbert}, {Lemaux}, {Moreau},
  {Paltani}, {Ribeiro}, {Salvato}, {Schaerer}, {Scodeggio}, {Sommariva},
  {Talia}, {Taniguchi}, {Tresse}, {Vergani}, {Wang}, {Charlot}, {Contini},
  {Fotopoulou}, {L{\'o}pez-Sanjuan}, {Mellier}, \& {Scoville}}]{VUDS}
{Le F{\`e}vre}, O., {Tasca}, L.~A.~M., {Cassata}, P., {et~al.} 2015, \aap, 576,
  A79

\bibitem[{{Le F{\`e}vre} {et~al.}(2005){Le F{\`e}vre}, {Vettolani}, {Garilli},
  {Tresse}, {Bottini}, {Le Brun}, {Maccagni}, {Picat}, {Scaramella},
  {Scodeggio}, {Zanichelli}, {Adami}, {Arnaboldi}, {Arnouts}, {Bardelli},
  {Bolzonella}, {Cappi}, {Charlot}, {Ciliegi}, {Contini}, {Foucaud},
  {Franzetti}, {Gavignaud}, {Guzzo}, {Ilbert}, {Iovino}, {McCracken}, {Marano},
  {Marinoni}, {Mathez}, {Mazure}, {Meneux}, {Merighi}, {Paltani}, {Pell{\`o}},
  {Pollo}, {Pozzetti}, {Radovich}, {Zamorani}, {Zucca}, {Bondi}, {Bongiorno},
  {Busarello}, {Lamareille}, {Mellier}, {Merluzzi}, {Ripepi}, \&
  {Rizzo}}]{vvds_main}
{Le F{\`e}vre}, O., {Vettolani}, G., {Garilli}, B., {et~al.} 2005, \aap, 439,
  845

\bibitem[{{Leja} {et~al.}(2019){Leja}, {Carnall}, {Johnson}, {Conroy}, \&
  {Speagle}}]{leja}
{Leja}, J., {Carnall}, A.~C., {Johnson}, B.~D., {Conroy}, C., \& {Speagle},
  J.~S. 2019, \apj, 876, 3

\bibitem[{{Lilly} {et~al.}(1995){Lilly}, {Le F{\`e}vre}, {Crampton}, {Hammer},
  \& {Tresse}}]{cfrs}
{Lilly}, S.~J., {Le F{\`e}vre}, O., {Crampton}, D., {Hammer}, F., \& {Tresse},
  L. 1995, \apj, 455, 50

\bibitem[{{Lilly} {et~al.}(2007){Lilly}, {Le F{\`e}vre}, {Renzini}, {Zamorani},
  {Scodeggio}, {Contini}, {Carollo}, {Hasinger}, {Kneib}, {Iovino}, {Le Brun},
  {Maier}, {Mainieri}, {Mignoli}, {Silverman}, {Tasca}, {Bolzonella},
  {Bongiorno}, {Bottini}, {Capak}, {Caputi}, {Cimatti}, {Cucciati}, {Daddi},
  {Feldmann}, {Franzetti}, {Garilli}, {Guzzo}, {Ilbert}, {Kampczyk}, {Kovac},
  {Lamareille}, {Leauthaud}, {Borgne}, {McCracken}, {Marinoni}, {Pello},
  {Ricciardelli}, {Scarlata}, {Vergani}, {Sanders}, {Schinnerer}, {Scoville},
  {Taniguchi}, {Arnouts}, {Aussel}, {Bardelli}, {Brusa}, {Cappi}, {Ciliegi},
  {Finoguenov}, {Foucaud}, {Franceschini}, {Halliday}, {Impey}, {Knobel},
  {Koekemoer}, {Kurk}, {Maccagni}, {Maddox}, {Marano}, {Marconi}, {Meneux},
  {Mobasher}, {Moreau}, {Peacock}, {Porciani}, {Pozzetti}, {Scaramella},
  {Schiminovich}, {Shopbell}, {Smail}, {Thompson}, {Tresse}, {Vettolani},
  {Zanichelli}, \& {Zucca}}]{zcosmos_main}
{Lilly}, S.~J., {Le F{\`e}vre}, O., {Renzini}, A., {et~al.} 2007, \apjs, 172,
  70

\bibitem[{{Magliocchetti} {et~al.}(2020){Magliocchetti}, {Pentericci},
  {Cirasuolo}, {Zamorani}, {Amorin}, {Bongiorno}, {Cimatti}, {Fontana},
  {Garilli}, {Gargiulo}, {Hathi}, {McLeod}, {McLure}, {Brusa}, {Saxena}, \&
  {Talia}}]{maglioc2020}
{Magliocchetti}, M., {Pentericci}, L., {Cirasuolo}, M., {et~al.} 2020, \mnras,
  493, 3838

\bibitem[{{Marchi} {et~al.}(2019){Marchi}, {Pentericci}, {Guaita}, {Talia},
  {Castellano}, {Hathi}, {Schaerer}, {Amorin}, {Bolzonella}, {Carnall},
  {Charlot}, {Chevallard}, {Cullen}, {Finkelstein}, {Fontana}, {Fontanot},
  {Garilli}, {Hibon}, {Koekemoer}, {Maccagni}, {McLure}, {Papovich},
  {Pozzetti}, \& {Saxena}}]{marchi2019}
{Marchi}, F., {Pentericci}, L., {Guaita}, L., {et~al.} 2019, \aap, 631, A19

\bibitem[{{Marigo} {et~al.}(2013){Marigo}, {Bressan}, {Nanni}, {Girardi}, \&
  {Pumo}}]{marigo}
{Marigo}, P., {Bressan}, A., {Nanni}, A., {Girardi}, L., \& {Pumo}, M.~L. 2013,
  \mnras, 434, 488

\bibitem[{{McLure} {et~al.}(2018){McLure}, {Pentericci}, {Cimatti}, {Dunlop},
  {Elbaz}, {Fontana}, {Nandra}, {Amorin}, {Bolzonella}, {Bongiorno}, {Carnall},
  {Castellano}, {Cirasuolo}, {Cucciati}, {Cullen}, {De Barros}, {Finkelstein},
  {Fontanot}, {Franzetti}, {Fumana}, {Gargiulo}, {Garilli}, {Guaita},
  {Hartley}, {Iovino}, {Jarvis}, {Juneau}, {Karman}, {Maccagni}, {Marchi},
  {M{\'a}rmol-Queralt{\'o}}, {Pompei}, {Pozzetti}, {Scodeggio}, {Sommariva},
  {Talia}, {Almaini}, {Balestra}, {Bardelli}, {Bell}, {Bourne}, {Bowler},
  {Brusa}, {Buitrago}, {Caputi}, {Cassata}, {Charlot}, {Citro}, {Cresci},
  {Cristiani}, {Curtis-Lake}, {Dickinson}, {Fazio}, {Ferguson}, {Fiore},
  {Franco}, {Fynbo}, {Galametz}, {Georgakakis}, {Giavalisco}, {Grazian},
  {Hathi}, {Jung}, {Kim}, {Koekemoer}, {Khusanova}, {Le F{\`e}vre}, {Lotz},
  {Mannucci}, {Maltby}, {Matsuoka}, {McLeod}, {Mendez-Hernandez},
  {Mendez-Abreu}, {Mignoli}, {Moresco}, {Mortlock}, {Nonino}, {Pannella},
  {Papovich}, {Popesso}, {Rosario}, {Salvato}, {Santini}, {Schaerer},
  {Schreiber}, {Stark}, {Tasca}, {Thomas}, {Treu}, {Vanzella}, {Wild},
  {Williams}, {Zamorani}, \& {Zucca}}]{mclure2018}
{McLure}, R.~J., {Pentericci}, L., {Cimatti}, A., {et~al.} 2018, \mnras, 479,
  25

\bibitem[{{Pannella} {et~al.}(2015){Pannella}, {Elbaz}, {Daddi}, {Dickinson},
  {Hwang}, {Schreiber}, {Strazzullo}, {Aussel}, {Bethermin}, {Buat},
  {Charmandaris}, {Cibinel}, {Juneau}, {Ivison}, {Le Borgne}, {Le Floc'h},
  {Leiton}, {Lin}, {Magdis}, {Morrison}, {Mullaney}, {Onodera}, {Renzini},
  {Salim}, {Sargent}, {Scott}, {Shu}, \& {Wang}}]{pannella}
{Pannella}, M., {Elbaz}, D., {Daddi}, E., {et~al.} 2015, \apj, 807, 141

\bibitem[{{Pentericci} {et~al.}(2018{\natexlab{a}}){Pentericci}, {McLure},
  {Franzetti}, {Garilli}, \& {the VANDELS team}}]{vandelsdr2}
{Pentericci}, L., {McLure}, R.~J., {Franzetti}, P., {Garilli}, B., \& {the
  VANDELS team}. 2018{\natexlab{a}}, arXiv e-prints, arXiv:1811.05298

\bibitem[{{Pentericci} {et~al.}(2018{\natexlab{b}}){Pentericci}, {McLure},
  {Garilli}, {Cucciati}, {Franzetti}, {Iovino}, {Amorin}, {Bolzonella},
  {Bongiorno}, {Carnall}, {Castellano}, {Cimatti}, {Cirasuolo}, {Cullen}, {De
  Barros}, {Dunlop}, {Elbaz}, {Finkelstein}, {Fontana}, {Fontanot}, {Fumana},
  {Gargiulo}, {Guaita}, {Hartley}, {Jarvis}, {Juneau}, {Karman}, {Maccagni},
  {Marchi}, {Marmol-Queralto}, {Nandra}, {Pompei}, {Pozzetti}, {Scodeggio},
  {Sommariva}, {Talia}, {Almaini}, {Balestra}, {Bardelli}, {Bell}, {Bourne},
  {Bowler}, {Brusa}, {Buitrago}, {Caputi}, {Cassata}, {Charlot}, {Citro},
  {Cresci}, {Cristiani}, {Curtis-Lake}, {Dickinson}, {Fazio}, {Ferguson},
  {Fiore}, {Franco}, {Fynbo}, {Galametz}, {Georgakakis}, {Giavalisco},
  {Grazian}, {Hathi}, {Jung}, {Kim}, {Koekemoer}, {Khusanova}, {Le F{\`e}vre},
  {Lotz}, {Mannucci}, {Maltby}, {Matsuoka}, {McLeod}, {Mendez-Hernandez},
  {Mendez-Abreu}, {Mignoli}, {Moresco}, {Mortlock}, {Nonino}, {Pannella},
  {Papovich}, {Popesso}, {Rosario}, {Salvato}, {Santini}, {Schaerer},
  {Schreiber}, {Stark}, {Tasca}, {Thomas}, {Treu}, {Vanzella}, {Wild},
  {Williams}, {Zamorani}, \& {Zucca}}]{vandelsdr1}
{Pentericci}, L., {McLure}, R.~J., {Garilli}, B., {et~al.} 2018{\natexlab{b}},
  \aap, 616, A174

\bibitem[{{Pentericci} {et~al.}(2018{\natexlab{c}}){Pentericci}, {Vanzella},
  {Castellano}, {Fontana}, {De Barros}, {Grazian}, {Marchi}, {Bradac},
  {Conselice}, {Cristiani}, {Dickinson}, {Finkelstein}, {Giallongo}, {Guaita},
  {Koekemoer}, {Maiolino}, {Santini}, \& {Tilvi}}]{candelsz7}
{Pentericci}, L., {Vanzella}, E., {Castellano}, M., {et~al.}
  2018{\natexlab{c}}, \aap, 619, A147

\bibitem[{{Rangel} {et~al.}(2013){Rangel}, {Nandra}, {Laird}, \&
  {Orange}}]{rangel}
{Rangel}, C., {Nandra}, K., {Laird}, E.~S., \& {Orange}, P. 2013, \mnras, 428,
  3089

\bibitem[{{S{\'a}nchez-Janssen} {et~al.}(2014){S{\'a}nchez-Janssen}, {Mieske},
  {Selman}, {Bristow}, {Hammersley}, {Hilker}, {Rejkuba}, \& {Wolff}}]{sanchez}
{S{\'a}nchez-Janssen}, R., {Mieske}, S., {Selman}, F., {et~al.} 2014, \aap,
  566, A2

\bibitem[{{Santini} {et~al.}(2015){Santini}, {Ferguson}, {Fontana}, {Mobasher},
  {Barro}, {Castellano}, {Finkelstein}, {Grazian}, {Hsu}, {Lee}, {Lee},
  {Pforr}, {Salvato}, {Wiklind}, {Wuyts}, {Almaini}, {Cooper}, {Galametz},
  {Weiner}, {Amorin}, {Boutsia}, {Conselice}, {Dahlen}, {Dickinson},
  {Giavalisco}, {Grogin}, {Guo}, {Hathi}, {Kocevski}, {Koekemoer},
  {Kurczynski}, {Merlin}, {Mortlock}, {Newman}, {Paris}, {Pentericci},
  {Simons}, \& {Willner}}]{Santini}
{Santini}, P., {Ferguson}, H.~C., {Fontana}, A., {et~al.} 2015, \apj, 801, 97

\bibitem[{{Saxena} {et~al.}(2020{\natexlab{a}}){Saxena}, {Pentericci},
  {Mirabelli}, {Schaerer}, {Schneider}, {Cullen}, {Amorin}, {Bolzonella},
  {Bongiorno}, {Carnall}, {Castellano}, {Cucciati}, {Fontana}, {Fynbo},
  {Garilli}, {Gargiulo}, {Guaita}, {Hathi}, {Hutchison}, {Koekemoer}, {Marchi},
  {McLeod}, {McLure}, {Papovich}, {Pozzetti}, {Talia}, \&
  {Zamorani}}]{saxena2020a}
{Saxena}, A., {Pentericci}, L., {Mirabelli}, M., {et~al.} 2020{\natexlab{a}},
  \aap, 636, A47

\bibitem[{{Saxena} {et~al.}(2020{\natexlab{b}}){Saxena}, {Pentericci},
  {Schaerer}, {Schneider}, {Amorin}, {Bongiorno}, {Calabr{\`o}}, {Castellano},
  {Cimatti}, {Cullen}, {Fontana}, {Fynbo}, {Hathi}, {McLeod}, {Talia}, \&
  {Zamorani}}]{saxena2020b}
{Saxena}, A., {Pentericci}, L., {Schaerer}, D., {et~al.} 2020{\natexlab{b}},
  \mnras, 496, 3796

\bibitem[{{Scodeggio} {et~al.}(2005){Scodeggio}, {Franzetti}, {Garilli},
  {Zanichelli}, {Paltani}, {Maccagni}, {Bottini}, {Le Brun}, {Contini},
  {Scaramella}, {Adami}, {Bardelli}, {Zucca}, {Tresse}, {Ilbert}, {Foucaud},
  {Iovino}, {Merighi}, {Zamorani}, {Gavignaud}, {Rizzo}, {McCracken}, {Le
  F{\`e}vre}, {Picat}, {Vettolani}, {Arnaboldi}, {Arnouts}, {Bolzonella},
  {Cappi}, {Charlot}, {Ciliegi}, {Guzzo}, {Marano}, {Marinoni}, {Mathez},
  {Mazure}, {Meneux}, {Pell{\`o}}, {Pollo}, {Pozzetti}, \& {Radovich}}]{vipgi}
{Scodeggio}, M., {Franzetti}, P., {Garilli}, B., {et~al.} 2005, \pasp, 117,
  1284

\bibitem[{{Scodeggio} {et~al.}(2018){Scodeggio}, {Guzzo}, {Garilli}, {Granett},
  {Bolzonella}, {de la Torre}, {Abbas}, {Adami}, {Arnouts}, {Bottini}, {Cappi},
  {Coupon}, {Cucciati}, {Davidzon}, {Franzetti}, {Fritz}, {Iovino}, {Krywult},
  {Le Brun}, {Le F{\`e}vre}, {Maccagni}, {Ma{\l}ek}, {Marchetti}, {Marulli},
  {Polletta}, {Pollo}, {Tasca}, {Tojeiro}, {Vergani}, {Zanichelli}, {Bel},
  {Branchini}, {De Lucia}, {Ilbert}, {McCracken}, {Moutard}, {Peacock},
  {Zamorani}, {Burden}, {Fumana}, {Jullo}, {Marinoni}, {Mellier}, {Moscardini},
  \& {Percival}}]{vipersfinal}
{Scodeggio}, M., {Guzzo}, L., {Garilli}, B., {et~al.} 2018, \aap, 609, A84

\bibitem[{{Speagle} {et~al.}(2014){Speagle}, {Steinhardt}, {Capak}, \&
  {Silverman}}]{speagle}
{Speagle}, J.~S., {Steinhardt}, C.~L., {Capak}, P.~L., \& {Silverman}, J.~D.
  2014, \apjs, 214, 15

\bibitem[{{Steidel} {et~al.}(2003){Steidel}, {Adelberger}, {Shapley},
  {Pettini}, {Dickinson}, \& {Giavalisco}}]{steidel2003}
{Steidel}, C.~C., {Adelberger}, K.~L., {Shapley}, A.~E., {et~al.} 2003, \apj,
  592, 728

\bibitem[{{Steidel} {et~al.}(2014){Steidel}, {Rudie}, {Strom}, {Pettini},
  {Reddy}, {Shapley}, {Trainor}, {Erb}, {Turner}, {Konidaris}, {Kulas}, {Mace},
  {Matthews}, \& {McLean}}]{kbss_mosfire}
{Steidel}, C.~C., {Rudie}, G.~C., {Strom}, A.~L., {et~al.} 2014, \apj, 795, 165

\bibitem[{{Steidel} {et~al.}(2004){Steidel}, {Shapley}, {Pettini},
  {Adelberger}, {Erb}, {Reddy}, \& {Hunt}}]{steidel2004}
{Steidel}, C.~C., {Shapley}, A.~E., {Pettini}, M., {et~al.} 2004, \apj, 604,
  534

\bibitem[{{Taylor}(2005)}]{topcat}
{Taylor}, M.~B. 2005, in Astronomical Society of the Pacific Conference Series,
  Vol. 347, Astronomical Data Analysis Software and Systems XIV, ed.
  P.~{Shopbell}, M.~{Britton}, \& R.~{Ebert}, 29

\bibitem[{{Thomas} {et~al.}(2020){Thomas}, {Pentericci}, {Le Fevre},
  {Zamorani}, {Schaerer}, {Amorin}, {Castellano}, {Carnall}, {Cristiani},
  {Cullen}, {Finkelstein}, {Fontanot}, {Guaita}, {Hibon}, {Hathi}, {Fynbo},
  {Khusanova}, {Koekemoer}, {McLeod}, {McLure}, {Marchi}, {Pozzetti}, {Saxena},
  {Talia}, \& {Bolzonella}}]{Thomas2020}
{Thomas}, R., {Pentericci}, L., {Le Fevre}, O., {et~al.} 2020, \aap, 634, A110

\bibitem[{{Turner} {et~al.}(2017){Turner}, {Cirasuolo}, {Harrison}, {McLure},
  {Dunlop}, {Swinbank}, {Johnson}, {Sobral}, {Matthee}, \& {Sharples}}]{kmos}
{Turner}, O.~J., {Cirasuolo}, M., {Harrison}, C.~M., {et~al.} 2017, \mnras,
  471, 1280

\bibitem[{{van der Wel} {et~al.}(2016){van der Wel}, {Noeske}, {Bezanson},
  {Pacifici}, {Gallazzi}, {Franx}, {Mu{\~n}oz-Mateos}, {Bell}, {Brammer},
  {Charlot}, {Chauk{\'e}}, {Labb{\'e}}, {Maseda}, {Muzzin}, {Rix}, {Sobral},
  {van de Sande}, {van Dokkum}, {Wild}, \& {Wolf}}]{legac}
{van der Wel}, A., {Noeske}, K., {Bezanson}, R., {et~al.} 2016, \apjs, 223, 29

\bibitem[{{Vettolani} {et~al.}(1997){Vettolani}, {Zucca}, {Zamorani}, {Cappi},
  {Merighi}, {Mignoli}, {Stirpe}, {MacGillivray}, {Collins}, {Balkowski},
  {Cayatte}, {Maurogordato}, {Proust}, {Chincarini}, {Guzzo}, {Maccagni},
  {Scaramella}, {Blanchard}, \& {Ramella}}]{esp}
{Vettolani}, G., {Zucca}, E., {Zamorani}, G., {et~al.} 1997, \aap, 325, 954

\bibitem[{{Williams} {et~al.}(2009){Williams}, {Quadri}, {Franx}, {van Dokkum},
  \& {Labb{\'e}}}]{williams}
{Williams}, R.~J., {Quadri}, R.~F., {Franx}, M., {van Dokkum}, P., \&
  {Labb{\'e}}, I. 2009, \apj, 691, 1879

\bibitem[{{Xue} {et~al.}(2011){Xue}, {Luo}, {Brandt}, {Bauer}, {Lehmer},
  {Broos}, {Schneider}, {Alexander}, {Brusa}, {Comastri}, {Fabian}, {Gilli},
  {Hasinger}, {Hornschemeier}, {Koekemoer}, {Liu}, {Mainieri}, {Paolillo},
  {Rafferty}, {Rosati}, {Shemmer}, {Silverman}, {Smail}, {Tozzi}, \&
  {Vignali}}]{xue}
{Xue}, Y.~Q., {Luo}, B., {Brandt}, W.~N., {et~al.} 2011, \apjs, 195, 10

\end{thebibliography}

\begin{appendix}
\section{Double measurements}
\label{appendix}
In Table \ref{zaccuracy}, we report the values of $\mathrm{n_{tot_{i,j}}}$ (the total number of pairs of measurements having  spectroscopic flags i and j), and of $\mathrm{n_{good_{i,j}}}$ (the number of  pairs of measurements having  spectroscopic flags i and j which are in agreement with each other), used to check the redshift probability as explained in Section \ref{accuracy}. In the table, each cell i,j reports the ratio $\mathrm{n_{good_{i,j}}}$/$\mathrm{n_{tot_{i,j}}}$ , where i (row number) is the flag associated to the measurement obtained with the shorter exposure time, and j (column number) is the flag associated with the measurement with the longer exposure time. For example, looking at the pairs for which the flag for the short exposure is 2 (row number) and for the long exposure is 3 (column number), we have 45 such double measurements, and 38 are in agreement, following the definition given in \ref{accuracy}.

\begin{table}
\caption{number of pairs of measurements used to check flag reliability}
\label{zaccuracy}
\centering
\begin{tabular}{|c |c|c|c|c|c| }
\hline
\diagbox{Flag\\t$\mathrm{_{short}}$}{Flag\\t$\mathrm{_{long}}$ }     & 1    & 2        & 3         &  4  & 9 \\
\hline     
1           &   6/30  &   10/41    &    14/28  &    2/5  &    1/ 4   \\
\hline
2 	    &     1/2   &   7/17    &    38/45  &  18/18  &    1/2 \\
\hline
3 	    &     0/1   &     1/1    &    29/33  &  48/51  &    1/1 \\
\hline
4 	    &     0/0   &     0/1    &      7/7  & 143/143 &    3/3 \\
\hline
9 	    &     0/2   &     0/1    &      9/9  &   10/10 &    10/12 \\
\hline
\end{tabular}
\end{table}

\end{appendix}

\end{document}